\documentclass[fleqn,usenatbib]{mnras}

\usepackage{newtxtext,newtxmath}

\usepackage[T1]{fontenc}
\usepackage{ae,aecompl}

\usepackage{graphicx}	
\usepackage{amsmath}	
\usepackage{amssymb}	
\usepackage{natbib}
\usepackage[]{threeparttable}
\usepackage{subfigure}

\usepackage{etoolbox}
\makeatletter
\patchcmd\@combinedblfloats{\box\@outputbox}{\unvbox\@outputbox}{}{\errmessage{\noexpand patch failed}}

\title[Reverberation Mapping of 3C~120]{Robotic reverberation mapping of the broad-line radio galaxy 3C~120}

\author[M. S. Hlabathe et al.]{
Michael Hlabathe$^{1, 2}$\thanks{E-mail: mh@saao.ac.za},
David Starkey$^{3}$,
Keith Horne$^{3}$,
Encarni Romero-Colmenero$^{2, 4}$,
\newauthor{
Steven Crawford$^{5}$,
Stefano Valenti$^{6, 7}$,
Hartmut Winkler$^8$,
Aaron Barth$^9$},
\newauthor{
Christopher Onken$^{10}$,
David Sand$^{11}$,
Tommaso Treu$^{12}$,
Aleksandar Diamond-Stanic$^{13}$},
\newauthor{
Carolin Villforth$^{14}$
}
\\
$^{1}$University of Cape Town, Private Bag X3, Rondebosch 7701, South Africa\\
$^{2}$South African Astronomical Observatory, P.O Box 9, Observatory 7935, Cape Town, South Africa\\
$^{3}$ SUPA Physics and Astronomy, University of St. Andrews, Fife, KY16 9SS, Scotland, UK\\
$^{4}$Southern African Large Telescope Foundation, P.O Box 9, Observatory 7935, Cape Town, South Africa\\
$^{5}$Space Telescope Science Institute, 3700 San Martin Drive, Baltimore, MD 21218, USA\\
$^{6}$Department of Physics, University of California, 1 Shields Avenue, Davis, CA 95616-5270, USA\\
$^{7}$Las Cumbres Observatory Global Telescope Network, 6740 Cortona Drive, Suite 102, Goleta, CA 93117, USA\\
$^{8}$Department of Physics, University of Johannesburg, P.O. Box 524, 2006 Auckland Park, South Africa\\
$^{9}$Department of Physics and Astronomy, University of California, 4129 Frederick Reines Hall, Irvine, CA, 92697-4575, USA\\
$^{10}$Research School of Astronomy and Astrophysics, Australian National University, Canberra ACT 2611, Australia\\
$^{11}$Department of Astronomy/Steward Observatory, 933 North Cherry Avenue, Room N204, Tucson, AZ 85721-0065, USA\\
$^{12}$Department of Physics and Astronomy, University of California, Los Angeles, CA 90095-1547, USA\\
$^{13}$Department of Physics and Astronomy, Bates College, 44 Campus Avenue, Lewiston ME 04240, USA\\
$^{14}$Department of Physics, University of Bath, Claverton Down Road, Bath BA2 7AY, UK
}
\date{Accepted XXX. Received YYY; in original form ZZZ}

\pubyear{2018}

\begin{document}
\label{firstpage}
\pagerange{\pageref{firstpage}--\pageref{lastpage}}
\maketitle

\begin{abstract}
We carried out photometric and spectroscopic observations of the well-studied broad-line radio galaxy 3C~120 with the Las Cumbres Observatory (LCO) global robotic telescope network from 2016 December to 2018 April as part of the LCO AGN Key Project on Reverberation Mapping of Accretion Flows. Here, we present both spectroscopic and photometric reverberation mapping results. We used the interpolated cross-correlation function (ICCF) to perform multiple-line lag measurements in 3C~120. We find the H$\gamma$, He II $\lambda 4686$, H$\beta$ and He~I $\lambda 5876$ lags of $\tau_{\text{cen}} = 18.8_{-1.0}^{+1.3}$, $2.7_{-0.8}^{+0.7}$, $21.2_{-1.0}^{+1.6}$, and $16.9_{-1.1}^{+0.9}$ days respectively, relative to the V-band continuum. Using the measured lag and rms velocity width of the H$\beta$ emission line, we determine the mass of the black hole for 3C~120 to be $M=\left(6.3^{+0.5}_{-0.3}\right)\times10^7\,(f/5.5)$~M$_\odot$. Our black hole mass measurement is consistent with similar previous studies on 3C~120, but with small uncertainties. In addition, velocity-resolved lags in 3C~120 show a symmetric pattern across the H$\beta$ line, 25 days at line centre decreasing to 17 days in the line wings at $\pm4000$~km~s$^{-1}$. We also investigate the inter-band continuum lags in 3C~120 and find that they are generally consistent with $\tau\propto\lambda^{4/3}$ as predicted from a geometrically-thin, optically-thick accretion disc. From the continuum lags, we measure the best fit value $\tau_{\rm 0} = 3.5\pm 0.2$ days at $\lambda_{\rm 0} = 5477\text{\AA}$. It implies a disc size a factor of $1.6$ times larger than prediction from the standard disc model with $L/L_{\rm Edd} = 0.4$. This is consistent with previous studies in which larger than expected disc sizes were measured. 
\end{abstract}




\section{Introduction}
It is believed that every active galactic nucleus (AGN) is powered by a supermassive black hole (BH) at its centre \citep{1971MNRAS.152..461L}. For a detailed description of AGN and their different components, otherwise known as the unified model of an AGN, see \cite{1993ARA&A..31..473A, 1995PASP..107..803U}. The supermassive BH in active galactic nuclei are believed to play a pivotal role in galaxy formation, thus encouraging efforts to understand their growth and distribution of mass over cosmological time.\\
\indent In particular, the observed scaling relationships between BH mass $M$ and host-galaxy properties, such as the velocity dispersion of the bulge stars $M-\sigma_{*}$ \citep{2000ApJ...539L...9F, 2000ApJ...539L..13G} or the bulge luminosity $M-L_{\text{bulge}}$ \citep{1995ARA&A..33..581K, 1998AJ....115.2285M}, indicate a strong connection between galaxy evolution and BH growth. To further investigate the nature of the BH-galaxy co-evolution, more BH masses must be accurately determined in both nearby galaxies and more distant AGNs. Dynamical methods based on high angular resolution kinematics of stars and gas have been widely used to measure masses of BHs in nearby galaxies \citep{2001AIPC..586..363K, 2005SSRv..116..523F}. However, they are limited to the spatially resolved kinematics of nearby (few tens of Mpc) galaxies even with the current technology. Yet, in Type 1 AGNs, the broad emission lines from the broad-line region (BLR) can be used to probe the inner structure of the AGN and therefore determine $M$ via a technique known as reverberation mapping \citep{1982ApJ...255..419B, 1993PASP..105..247P, 2001sac..conf....3P}.\\
\indent Reverberation mapping (RM) measures the mean time delay or lag $\tau$ between changes in the continuum from the accretion disc and corresponding changes in the broad emission lines, thus allowing for the direct measurement of the BLR size $(R_{BLR} = c\tau)$ assuming light-travel time effects. Typically, the time delay is found using traditional cross-correlation methods, such as the interpolation cross-correlation function \citep[ICCF,][]{1986ApJ...305..175G, 1987ApJS...65....1G, 1994PASP..106..879W} or the discrete correlation function \citep[DCF,][]{1988ApJ...333..646E}. Over time, other approaches have been introduced for lag determination, such as Stochastic Process Estimation for AGN Reverberation \citep[SPEAR and/or python implementation JAVELIN,][]{2011ApJ...735...80Z} and quite recently the Continuum REprocessed AGN Markov Chain \citep[CREAM,][]{2016MNRAS.456.1960S, 2017ApJ...835...65S}, which have been found to perform better statistically on data with seasonal gaps or irregular sampling than the traditional methods \citep{2017ApJ...851...21G, 2017ApJ...846...79L}.\\
\indent By combining the measured $\tau$ and the velocity width $\Delta V$ (either from FWHM or line dispersion $\sigma_{\text{line}}$) of the broad emission line, and assuming virialized motions of the gas within the BLR, we can estimate the central black hole mass as
\begin{equation}
M = f\,\dfrac{c\,\tau\,\Delta V^{2}}{G}\,,
\end{equation}
\label{BH_mass_formula}
\noindent where $\tau$ is the measured lag of the broad emission line, $c$ is the speed of light, $G$ is the gravitational constant, and $f$ is a dimensionless factor which takes the structure, kinematics and orientation effects of the BLR into account.\\
\indent RM has been applied successfully in multiple spectroscopic campaigns, yielding > 50 BH masses to date \citep[e.g.][]{1999AN....320..319W, 2000ApJ...533..631K, 2004ApJ...613..682P, 2009ApJ...705..199B, 2009ApJ...692..246D, 2011ApJ...743L...4B, 2015ApJS..217...26B, 2015ApJ...813L..36V, 2016ApJ...818...30S, 2017ApJ...840...97F, 2018ApJ...868...76G, 2018ApJ...856....6D}, see also \cite{2015PASP..127...67B}. Although RM can be used to find BH masses in nearby and distant AGNs, it does, however, require long-term monitoring, high cadence and signal-to-noise, which is observationally hard to achieve using ground-based telescopes. Fortunately, RM also provides a scaling relationship between BLR size and the AGN luminosity at 5100\AA~($R_{BLR}-L_{5100}$) \citep[e.g.,][]{2013ApJ...767..149B}, thus allowing for the measurement of BH mass using single-epoch spectrum of the AGN \citep[e.g.,][]{1999AN....320..319W, 2002ApJ...571..733V}.\\
\indent RM can also be used to provide some insight into the structure and size of the accretion disc in AGN using the wavelength-dependent continuum delays, wherein ionizing photons from a compact region near BH propagate outwards and is reprocessed into UV/Optical continuum emission with light-travel time delays that increase with wavelength \citep[e.g.][]{2007MNRAS.380..669C}. By measuring the continuum time delay $\tau$ as a function of wavelength $\lambda$, RM can probe the radial temperature profile of the accretion disc and test the $\tau\propto\lambda^{4/3}$ prediction of the standard thin-disc model of \cite{1973A&A....24..337S}.\\
\indent Here, we undertake a robotic RM study of 3C~120 (Table~\ref{object_properties}) from long-term monitoring with the Las Cumbres Observatory (LCO) global robotic telescope network. 3C~120 was shown to exhibit a type 1 Seyfert spectrum by \citet{1967PASP...79..369S} and it was one of the first radio sources for which variability was confirmed \citep{1966ApJ...146..634P}. Furthermore, 3C~120 became the subject of much interest when the apparent superluminal motion of its radio jet was discovered \citep{1997ApJ...488..675W, 2011ApJ...733...11G}.

3C~120 has also been the subject of several previous reverberation studies over the years. Observational campaigns in the late 1990's \citep{1998ApJ...501...82P}, later reanalysed by \cite{2004ApJ...613..682P} determined a time lag of $\sim 40$ days between the continuum and the H$\beta$ line. \citet{2014A&A...566A.106K} performed a reverberation analysis of a set of 3C~120 optical spectra secured between 2008 September and 2009 March, concluding that the H$\beta$ lag relative to the continuum is $27.9_{-5.9}^{+7.1}$ days. \citet{2018ApJ...869..142D} determined the H$\beta$-to-continuum delay to be $20.2_{-4.2}^{+5.0}$ days. 3C~120 has also been the subject of infrared dust reverberation mapping by \cite{2018A&A...620A.137R}. Using our high signal-to-noise observations, we aim to accurately determine $M$ in 3C~120 along with measuring the inter-band continuum lags.  

We arranged this paper as follows: in Section~\ref{observations_and_data_reduction} we describe the observations and data reduction. Section~\ref{time_series_analysis} reviews the time-series analysis methodology. Section \ref{continuum_lags_accretion_disc} presents the inter-band continuum delays in context with the standard accretion disc theory. Section~\ref{black_hole_mass} presents the black hole mass measurements. In Section~\ref{velocity_resolved_lag_measurements}, we investigate the structure and kinematics of the BLR in 3C~120. We discuss our results and summarize our conclusions in Sections~\ref{discussion} and \ref{conclusion} respectively. We assumed a cosmological model with $H_{0} = 73.0$ km s$^{-1}$ Mpc$^{-1}$, $\Omega_{m} = 0.27$ and $\Omega_{\Lambda} = 0.73$.

\section{Observations and Data Reduction}
\label{observations_and_data_reduction}
To carry out a reverberation mapping analysis on 3C~120, we require continuum light curve(s) which we measure from photometry and emission-line light curve(s) from spectroscopy. The data employed here were acquired as part of the Las Cumbres Observatory (LCO) AGN Key Project titled Reverberation Mapping of AGN Accretion Flows \citep{2015ApJ...813L..36V}. The observations include 6-band photometric imaging
(Section~\ref{photometry}) and FLOYDS spectroscopy (Section~\ref{spectroscopy}).
\begin{table}
\caption{Object properties}
\label{object_properties}
\begin{threeparttable}
\begin{tabular}{ccccc}
\hline \hline
Object & RA$^{(1)}$ & DEC$^{(1)}$ & $D_{\rm L}^{(2)}$ & $z^{(1)}$\\
       & (J2000) & (J2000) & (Mpc) & \\
\hline
3C~120 & 04 33 11.1 & +05 21 16 & 139 & 0.0330\\
\hline
\end{tabular}
\begin{tablenotes}
\item[1] Values from NED database.
\item[2] Adopted luminosity distance for our cosmological model.
\end{tablenotes}
\end{threeparttable}
\end{table}
\subsection{Photometry}
\label{photometry}
\begin{figure*}
\centering
\includegraphics[scale = 0.55]{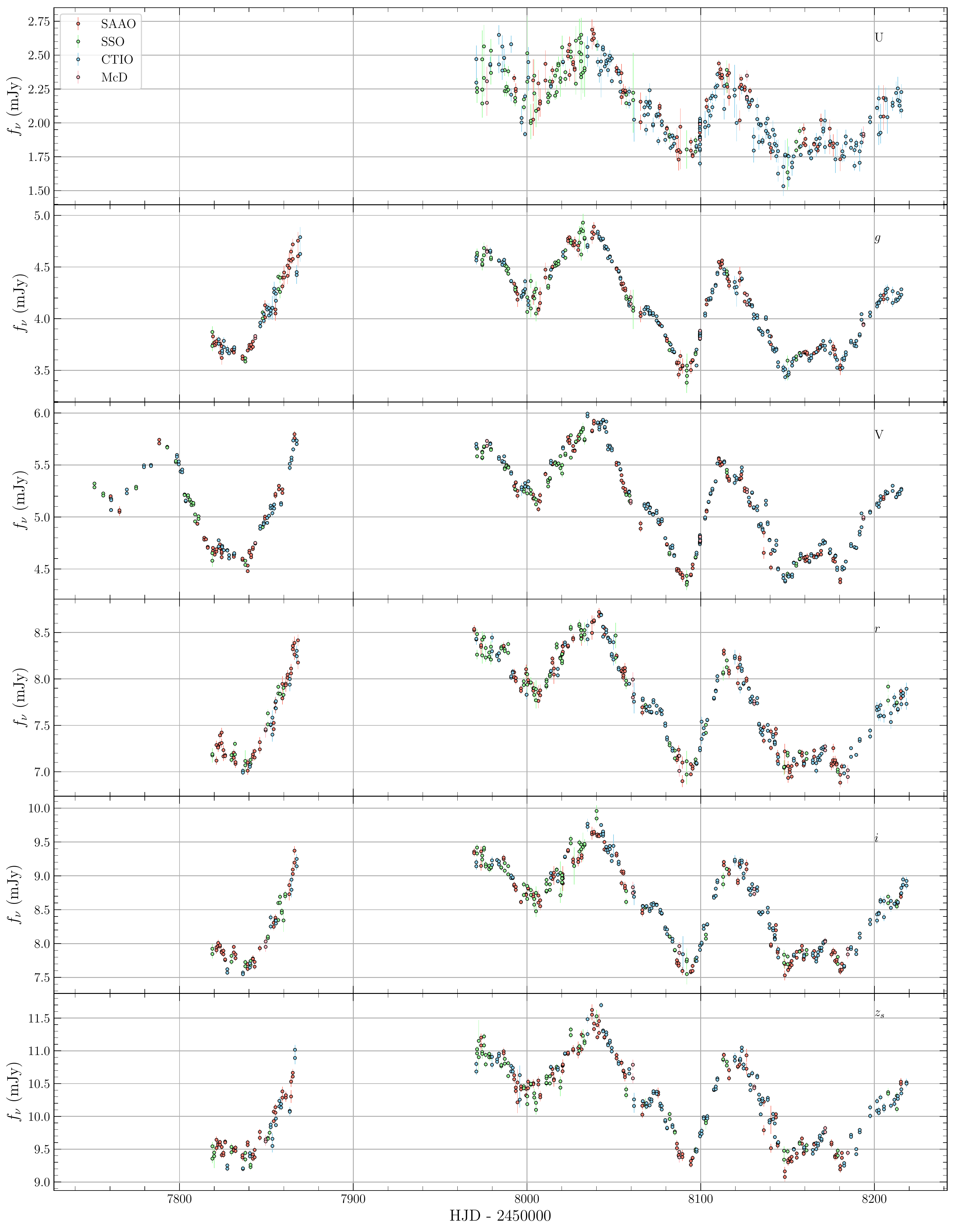}
\caption{The 3C~120 continuum light curves in (top to bottom) U, $g'$, V, $r'$, $i'$, $z_s$. Data are ordered by wavelength.}
\label{cont_lcurves}
\end{figure*}

Our photometric monitoring campaign of 3C~120 in 6 bands (U, $g'$, V, $r'$, $i'$, $z_s$) ran from 2016 December to 2018 April using the LCO 1m robotic telescopes \citep{2013PASP..125.1031B} deployed at Siding Spring Observatory (SSO), South African Astronomical Observatory (SAAO), McDonald Observatory (McD) and Cerro Tololo Interamerican Observatory (CTIO). A journal of photometric observations is presented in Table~\ref{summary_of_observations}, and the resulting light curves in Figure~\ref{cont_lcurves}. All observations were taken with Sinistro, a 4k$\times$4k-pixel quad-readout CCD camera with a $26.5\times26.5$ arcmin field of view and a pixel scale of 0.389 arcsec/pixel. We obtained 563 observations in the Johnson V filter, 432 in the Johnson U filter, 508 in the SDSS $g'$ filter, 498 in the SDSS $r'$ filter, 480 in the SDSS $i'$ filter and 460 in the Pan-STARRS $z_s$ filter with nearly daily cadence during 50 nights before and 250 nights after the 100-day solar conjunction gap.
\begin{table*}
\caption{Summary of photometric observations for 3C~120.}
\label{summary_of_observations}
\begin{threeparttable}
\begin{tabular}{ccccccc}
\hline \hline
Filter$^{(1)}$ & $\lambda_{\text{pivot}}^{(2)}$ & LCO site$^{(3)}$ & Date range$^{(4)}$ & $t_{\rm exp}$~(s) & Epochs$^{(5)}$ & Total$^{(6)}$\\
\hline
U & $3656\text{\AA}$ & SAAO & 20170826-20180327 & 300 & 107 & 432\\
  &	& McD & 20170810-20180315 & & 14 &\\
  &	& SSO & 20170806-20180207 & & 90 &\\
  &	& CTIO & 20170804-20180406 & & 221 &\\
\hline
$g'$ & $4770\text{\AA}$ & SAAO & 20170306-20180327 & 120 & 149 & 508\\
  &	& McD & 20170330-20180315 & & 14 &\\
  &	& SSO & 20170306-20180207 & & 91 &\\
  &	& CTIO & 20170309-20180406 & & 254 &\\
\hline
V & $5477\text{\AA}$ & SAAO & 20170106-20180327 & 300 & 156 & 563\\
  &	& McD & 20170330-20180315 & & 14 &\\
  &	& SSO & 20161228-20180207 & & 113 &\\
  &	& CTIO & 20170106-20180406 & & 280 &\\
\hline
$r'$ & $6231\text{\AA}$ & SAAO & 20170308-20180406 & 120 & 171 & 498\\
  &	& McD & 20170405-20180306 & & 12 &\\
  &	& SSO & 20170306-20180404 & & 125 &\\
  &	& CTIO & 20170314-20180409 & & 190 &\\
\hline
$i'$ & $7625\text{\AA}$ & SAAO & 20170308-20180406 & 120 & 158 & 480\\
  &	& McD & 20170405-20180306 & & 12 &\\
  &	& SSO & 20170306-20180404 & & 124 &\\
  &	& CTIO & 20170314-20180309 & & 186 &\\
\hline
$z_s$ & $8660\text{\AA}$ & SAAO & 20170308-20180406 & 240 & 160 & 460\\
   & & McD & 20170405-20180306 & & 11 &\\
   & & SSO & 20170306-20180304 & & 106 &\\
   & & CTIO & 20170314-20180409 & & 183 &\\
\hline
\end{tabular}
\begin{tablenotes}
\item[1] The filter through which the data were taken.
\item[2] Filter centroid wavelength.
\item[3] SAAO (South African Astronomical Observatory), McD (McDonald Observatory), SSO (Siding Spring Observatory), CTIO (Cerro Tololo Interamerican Observatory).
\item[4] Start and end date of observations, we note the $\sim 100$~day gap due to solar conjunction.
\item[5] The number of observations taken in that date range per filter.
\item[6] The total number of observations taken in that filter from all telescopes combined.
\end{tablenotes}
\end{threeparttable}
\end{table*}

\indent  The basic data reduction steps were carried out using the LCO BANZAI data pipeline \citep{2013PASP..125.1031B}, which performs flat fielding and bias subtraction. Subsequent to this, all images were cleaned of cosmic rays applying the L.A.Cosmic algorithm \citep{2001PASP..113.1420V}.\\
\indent To construct differential light curves in the respective bands, we performed a multi-aperture photometry in AstroImageJ. AstroImageJ is an image analysis tool designed to automatically go through a set of time-series images and measure light curves \citep{2017AJ....153...77C}.
\begin{figure}
\centering
\includegraphics[width = \columnwidth]{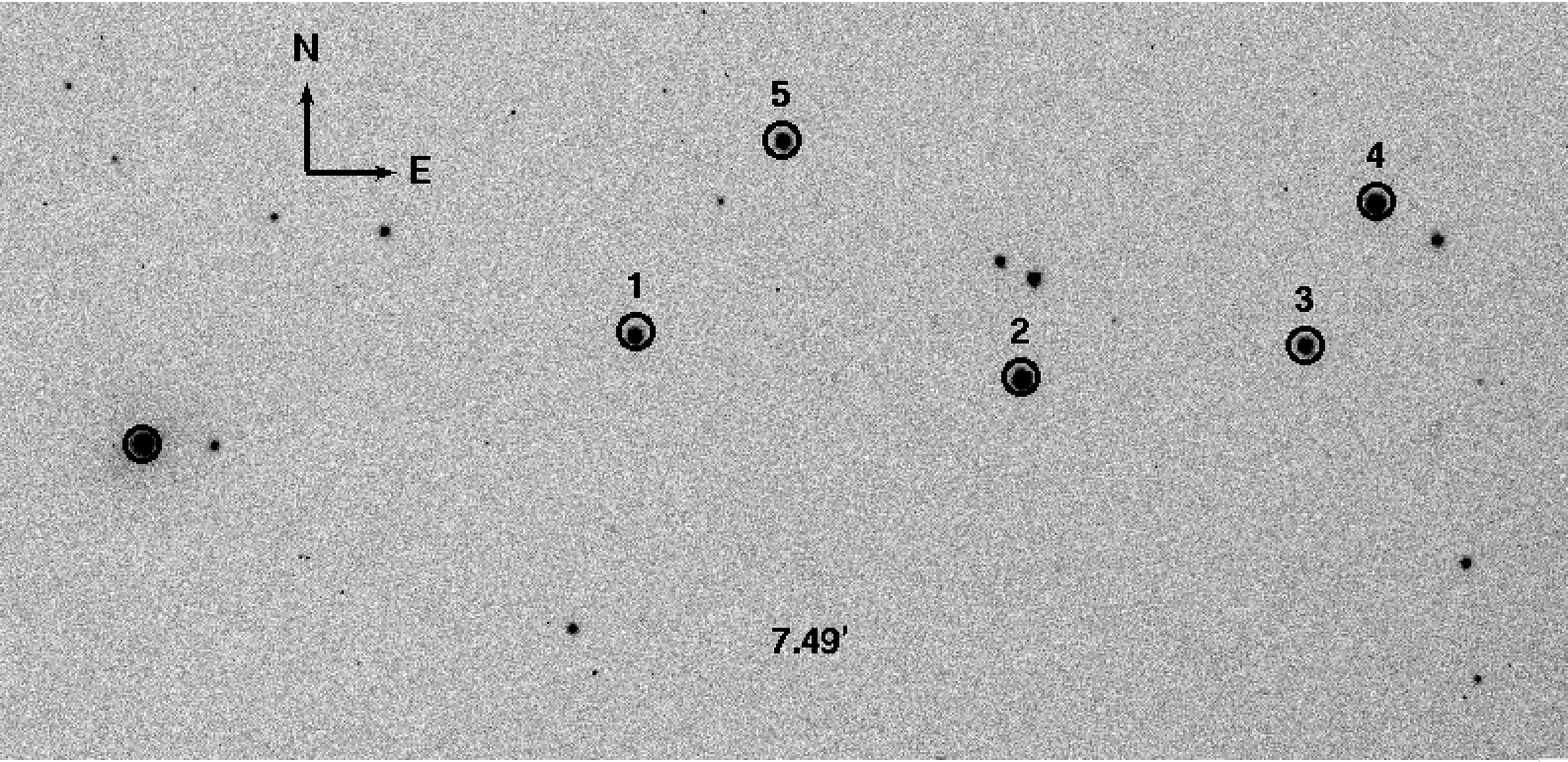}
\caption{A cut-out of the full Sinistro V-band image of 3C~120 taken with CTIO 1m telescope. The comparison stars used in the differential photometry are indicated. Image is 7.49 arcmin across, the N and E pointers/arrows are 0.3 arcmin in length.}
\label{comp_stars}
\end{figure}

\indent We analyzed the dataset from each filter separately but in the same manner. Firstly, we aligned the V images using \textit{Align stack using WCS} functionality in AstroImageJ. Using visual inspection, we selected from a set of aligned images, one of the images with the best seeing. We selected 5 comparison stars on the basis of their brightness and relative isolation (Figure \ref{comp_stars}). We then measured the FWHM in pixels of those stars using \textit{Plot seeing profile} in AstroImageJ, and recorded the measurements. Next, we randomly selected 20 more images from the set and measured the FWHM of the comparison stars. We then combined all the measured FWHM to produce the average. Using the average value, an aperture radius and sky annulus radii were then determined as recommended in \cite{2017AJ....153...77C}. This resulted in an aperture radius of 12 pixels (5.0 arcsec) and an inner and outer sky aperture radii of 22 and 33 pixels respectively. We adopted the same aperture and sky annuli radii for all filters.\\

\indent AstroImageJ produces a measurements table which includes among other parameters, sky-subtracted counts of the target and that of the chosen comparison stars. To perform differential photometry, we measure instrumental magnitudes of the target and comparison stars from the measured counts. We then selected a reference frame (first image in the set) as the zero point. We calculated the magnitude difference $(\Delta m)$ of each comparison star between the reference epoch and each epoch. We also calculated the rms variability of each comparison star based on the resulting $(\Delta m)$. The comparison star with lowest rms variability was selected as the main comparison star and others as check stars. For each filter, a different comparison star was selected as indicated in Table \ref{absolute_calibration_info}. We subtracted the main comparison star from the check stars $(\Delta m - \Delta m_{\rm comp})$. As a consistency check, we found the check stars to yield rms intrinsic variations of about 1\%. We then also subtracted the main comparison star from the target to get the differential magnitudes of the target ($m_{\rm target} - m_{\rm comp}$). We calibrated the differential magnitudes of the target using the catalog \citep[APASS,][]{2015AAS...22533616H} and \citep[Pan-STARRS,][]{2016arXiv161205560C} magnitudes of the comparison stars. As a reference U-band magnitude could not be found for the comparison stars, we estimate the U-band flux for the comparison stars based on the B-V, U-B colors \citep{1970A&A.....4..234F, 2001ApJ...558..309D} and assuming a 0.2 mag uncertainty on the resulting U magnitude. We then performed a conversion to AB system for Johnson U and V magnitudes \citep{2007AJ....133..734B}. Finally, we flux-calibrated the differential light curve of the target to mJy. The same procedure was repeated for each filter. In Table \ref{absolute_calibration_info}, we give the adopted mean AB magnitude and its uncertainty for each comparison star.\\
\begin{table*}
\caption{The comparison stars AB magnitudes for each filter.}
\label{absolute_calibration_info}
\begin{threeparttable}
\begin{tabular}{cccccc}
\hline \hline
Filter & Comp 1 & Comp 2 & Comp 3 & Comp 4 & Comp 5\\
\hline
U & $16.5\pm 0.2^{*}$ & $16.3\pm 0.2$ & $17.5\pm 0.2$ & $16.8\pm 0.2$ & $17.6\pm 0.2$\\
$g'$ & $15.737\pm 0.076^{*}$ & $14.994\pm 0.024$ & $15.881\pm 0.081$ & $15.071\pm 0.05$ & $15.918\pm0.049$\\
V & $15.388\pm 0.058$ & $14.480\pm 0.019^{*}$ & $15.417\pm 0.035$ & $14.449\pm 0.044$ & $15.424\pm 0.031$\\
$r'$ & $15.182\pm 0.047^{*}$ & $14.162\pm 0.063$ & $14.962\pm 0.077$ & $14.014\pm 0.041$ & $15.110\pm 0.042$\\
$i'$ & $14.89\pm 0.06$ & $13.86\pm 0.04$ & $14.58\pm 0.04^{*}$ & $13.54\pm 0.06$ & $14.8\pm0.1$\\
$z_s$ & & $13.7374\pm 0.0033^{*}$ & $14.4569\pm 0.0044$ & $13.3687\pm 0.0005$ & $14.6546\pm0.0057$\\
\hline
\end{tabular}
\begin{tablenotes}
\item[*] Comparison stars used in the calibration.
\end{tablenotes}
\end{threeparttable}
\end{table*}
\indent In the resulting light curves, we removed extreme outliers (verified by inspection of the images to be caused by, for example, cosmic ray hits, bad ccd pixels, strong sky gradients, clouds, etc). This left us with 541(96\%) data points on the final light curve in V, 380(88\%) in U, 476(94\%) in $g'$, 446(90\%) in $r'$, 439(92\%) in $i'$ and 425(92\%) in $z_s$. We present the resulting 3C~120 light curves in each band in Figure \ref{cont_lcurves}. The light curves exhibit strong variations with similar structure across all the photometric bands.
Several 10-20\% features on 30-100 day timescales
are detected with high S/N, as well as numerous smaller features. These well-detected variations provide the basis for measuring inter-band time delays. We also note that there may be potential emission-line contamination of the broad-band photometric light curves, which could later impact on the measured time lags between the bands.
\subsection{Spectroscopy}
\label{spectroscopy}
The spectroscopic observations were acquired robotically with the FLOYDS spectrographs between 2017 March and 2018 March. The FLOYDS spectrographs are two almost identical spectrographs on the LCO 2m telescopes, the Faulkes Telescope North (FTN) at Haleakela, Hawaii, and South (FTS) at Siding Springs in Australia. The spectrographs cover concurrently a wavelength range of $4800 - 10000\text{\AA}$ in first order and $3200 - 5900\text{\AA}$ in second order. The pixel scales are 3.51 \AA/pixel in first order and 1.74 \AA/pixel in second order, which corresponds to resolutions of $17\text{\AA}$ and $9\text{\AA}$ using a slit width of 2 arcseconds, respectively. For brevity, these will be referred to as the red and blue side respectively. We acquired a total of 149 spectroscopic epochs at a median sampling of 2 days. Spectra were taken in pairs with $1800$~s per spectrum.\\
\indent The data were reduced separately for the blue and red sides using the AGN FLOYDS pipeline written in PyRAF/Python\footnote{https://github.com/svalenti/FLOYDS\textunderscore pipeline.git}. The pipeline includes overscan subtraction, flat fielding, defringing, cosmic-ray rejection using the L.A.Cosmic algorithm \citep{2001PASP..113.1420V}, order rectification, spectral extraction, flux and wavelength calibration. All these steps were performed without human interaction.\\
\indent We inspected the resulting 1-D spectra visually and we discarded the bad spectra, either due to significantly poor signal to noise or the pipeline failing to extract the 1-D spectra. Doing this procedure, we ended up with 128 blue spectra (97 FTN, 31 FTS) and 137 red spectra (98 FTN, 39 FTS). Since the H$\beta$ emission line is observed in both orders (red and blue), it permits for simultaneous measurements of the line providing a way to cross check our measurements \citep{2015ApJ...813L..36V}.\\
\indent We analysed separately the blue and red spectra, and further divided the FTN and FTS spectra, which have slightly different wavelength scales and flux calibrations. We also exclude the H$\alpha$ line from further analysis as it is affected by fringing.
\subsection{Spectral modeling with PrepSpec}
\label{prepspec}
\begin{figure*}
\centering
\subfigure[]{\includegraphics[scale=0.48]{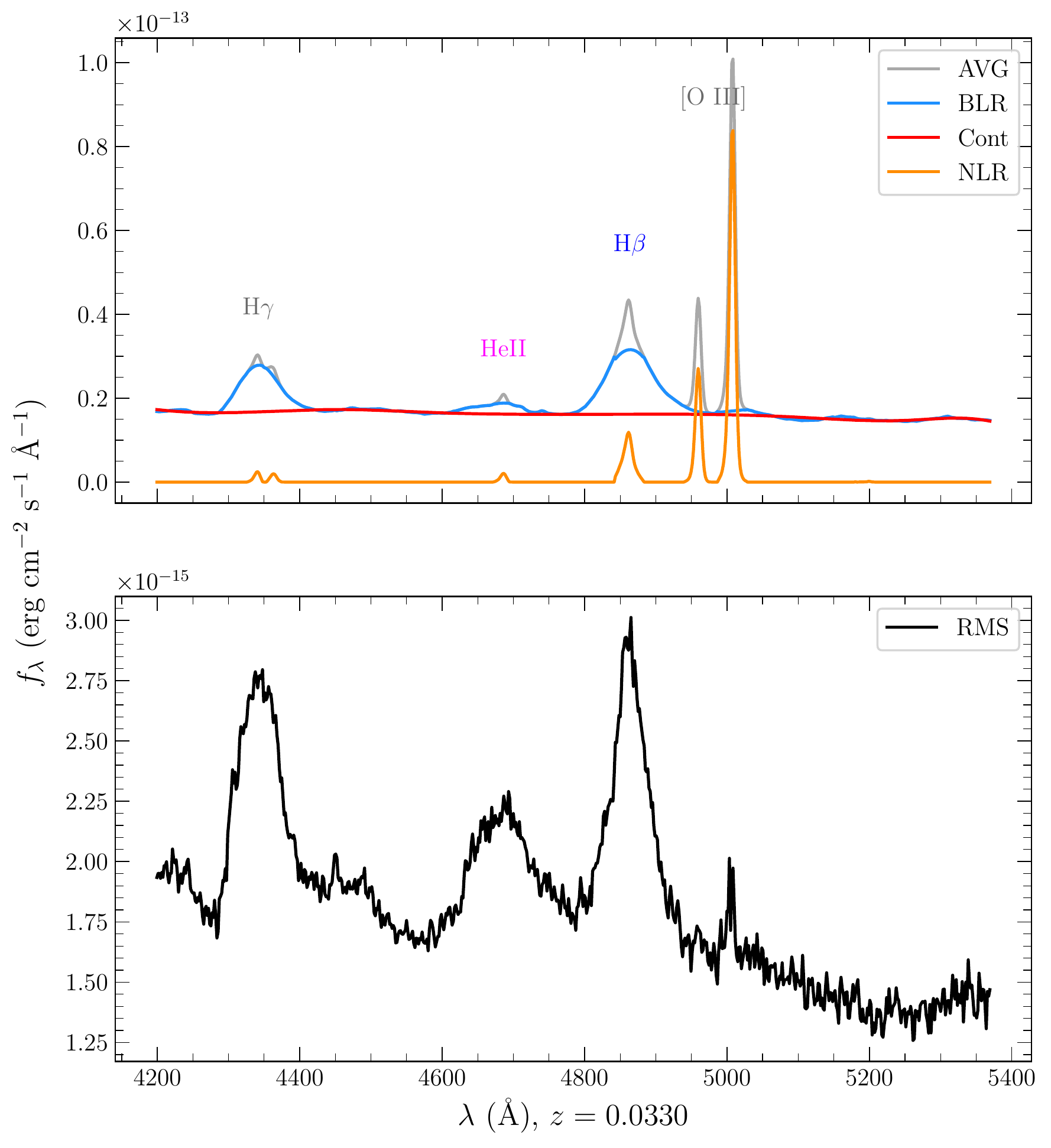}}
\subfigure[]{\includegraphics[scale=0.48]{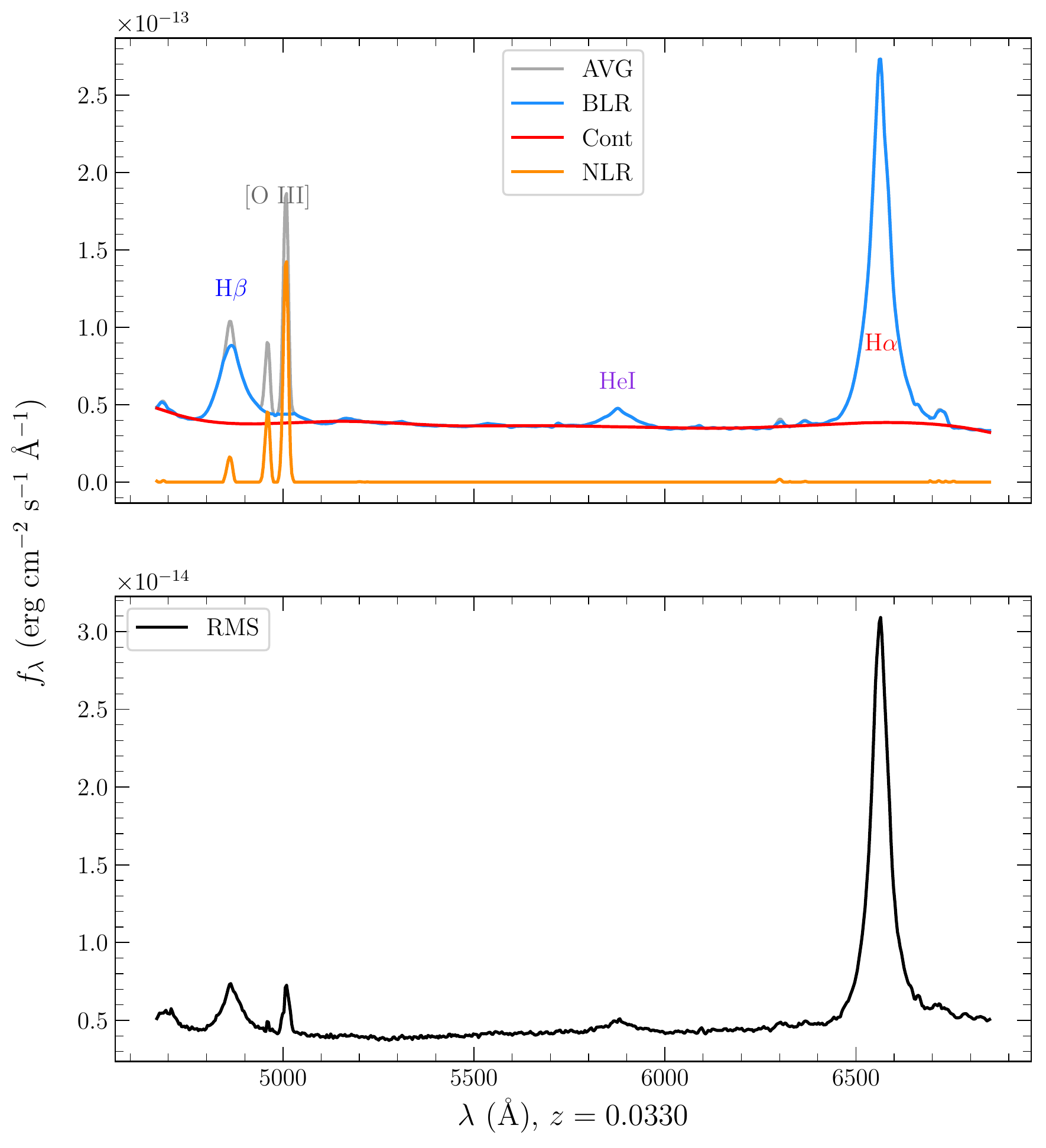}}
\caption{(Left) Mean (top) and rms (bottom) spectrum of 3C~120 from FTN data (blue-side), which includes 97 epochs. Labeled are H$\gamma$ $\lambda 4340$, He~II $\lambda 4686$, H$\beta$ $\lambda 4861$ and [O III] $\lambda 4959,5007$ emission-lines. (Right) Mean (top) and rms (bottom) spectrum of 3C~120 from FTN data (red-side), which includes 98 epochs. Labeled are H$\beta$ $\lambda 4861$, [O III] $\lambda 4959,5007$, and He~I $\lambda 5876$ emission-lines. The H$\alpha$ line is excluded from our time-series analysis due to it being affected by fringing.}
\label{mean_specs}
\end{figure*}

\begin{figure*}
\centering
\includegraphics[width = \textwidth]{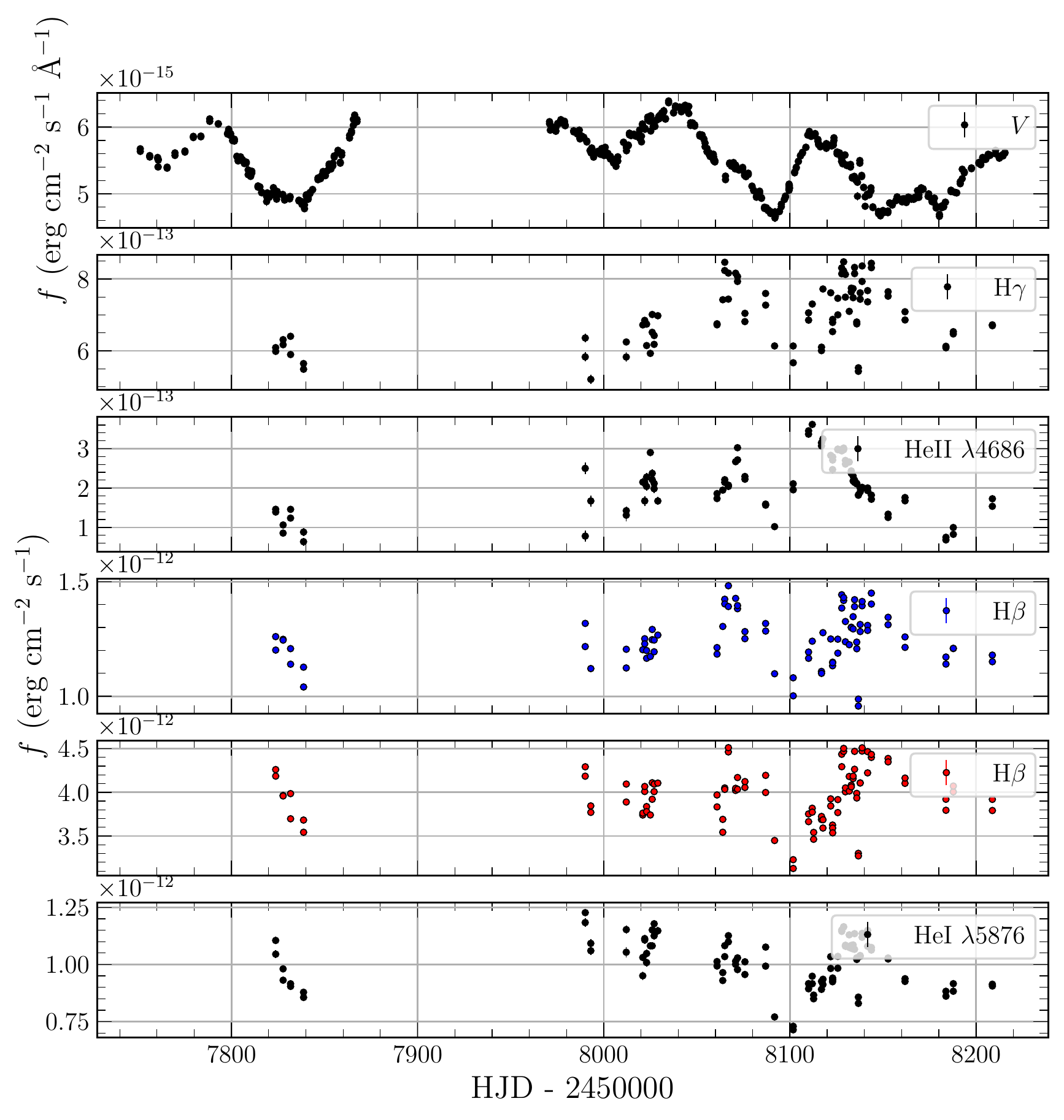}
\caption{V-band continuum and broad emission-line (H$\gamma$ $\lambda 4340$, He~II $\lambda 4686$, H$\beta$ $\lambda 4861$, He~I $\lambda 5876$) light curves from PrepSpec using FTN data.}
\label{line_lcurves}
\end{figure*}

To extract the emission-line light curves, we employ a multi-epoch spectral fitting tool called PrepSpec \citep{2015ApJS..216....4S, 2016ApJ...818...30S}. PrepSpec was written in Fortran by K. Horne and is optimized to perform simultaneous spectral fitting of AGN spectra. We analyse spectra from the blue and red side separately. For a full description of PrepSpec, see \cite{2016ApJ...818...30S}.\\
\indent To mitigate flux calibration errors due to slit losses and changes in atmospheric transparency, PrepSpec fits a model to the input spectra defined as
\begin{equation}
\mu(t, \lambda) = p(t)[A(\lambda) + B(t, \lambda) + C(t, \lambda)]\,, 
\end{equation}
where $A(\lambda)$ is the average spectrum, $B(t, \lambda)$ is the variable BLR and $C(t, \lambda)$ is the variable continuum. $p(t)$ are the time-dependent photometric corrections determined by assuming that the narrow emission-line ([O III]$\lambda 4959,5007$ and other) fluxes remain constant throughout the RM monitoring period. Even though narrow emission-lines have been found to vary over long timescales \citep[$\sim$several years,][]{2013ApJ...779..109P}, this assumption is valid for short timescales (days to months). The variable BLR is modeled as a sum over lines of the line profile multiplied by the line light curve,
\begin{equation}
B(t, \lambda) = \sum_{\ell=1}^{N_\ell}B_{\ell}(\lambda)L_{\ell}(t).
\end{equation}
Here $\langle L \rangle = 0$ and $\langle L^2 \rangle = 1$ to resolve the degeneracy between $B_{\ell}(\lambda)$ and $L_{\ell}(t)$, so that the rms profile of the line $\ell$ is $B_{\ell}(\lambda)$. The light curve of the line $\ell$ is $L_{\ell}(t)$ multiplied by the integral of the line flux in the rms spectrum
\begin{equation}
B_{\ell}(t) = L_{\ell}(t)\int B_{\ell}(\lambda)\,d\lambda\,. 
\end{equation}
The average spectrum $A(\lambda)$ is the sum of the broad-line, continuum and narrow-line components. Mathematically,
\begin{equation}
A(\lambda) = \bar{F}(\lambda) + N(\lambda)\,.
\end{equation}
In this case $\bar{F}(\lambda)$ encapsulates both the broad-line and continuum components, and $N(\lambda)$ is the narrow-line component. Unlike many RM studies in which a power-law is used to model the changes in the continuum, PrepSpec uses a low-order polynomial in $\log{\lambda}$ with time-dependent coefficients to model the variations in the continuum $C(t, \lambda)$.
PrepSpec also includes small corrections taking into account spectral blurring due to seeing and small wavelength shifts due to instrument flexure and differential refraction.\\
\indent PrepSpec optimises the model parameters to fit all spectra simultaneously. PrepSpec outputs the mean and root-mean-square (rms) residual line profiles (Figure~\ref{mean_specs}). The mean spectra exhibit strong narrow [OIII] emission-lines, with weaker narrow-line components on the H$\gamma$, H$\beta$, He~II $\lambda 4686$, He I $\lambda 5876$ lines. The PrepSpec model uses all of these narrow features to define the calibration tweaks. The blue and red spectra both cover H$\beta$ and [OIII], providing two independent measures of the H$\beta$ line profile and variations. The continuum is roughly flat in the mean $f_\lambda$ spectrum and significantly bluer in the rms spectrum. The broad emission-lines are more prominent in the rms than in the mean spectrum, particularly for He~II $\lambda 4686$.\\
\indent PrepSpec also produces measurements of the integrated broad emission-line fluxes (Figure~\ref{line_lcurves}), linewidth measurements (FWHM, $\sigma_{\rm line}$) and their statistical errors for each of the broad-lines. The measured linewidths are listed in Table \ref{line_widths} for comparison. The highly ionized He~II $\lambda 4686$ line exhibits broader linewidth in the rms spectrum due to its close origin to the ionizing source. Generally, we measure small uncertainties on the linewidths due to high S/N of our data. However, we do note that because of the ambiguity in separating the NLR and BLR components, and placing the continuum, the measured linewidths may be over/under estimated.\\
\indent Correlated variations across all the emission-lines are seen in Figure~\ref{line_lcurves}. The He~II $\lambda 4686$ variations track the V-band variations quite closely, while He~I $\lambda 5876$ and the Balmer line variations exhibit evidence for a more substantial lag, despite their relatively large scatter around HJD-2450000 = 8100-8150. For example, the V-band light curve rises from 8090-8110 and then falls until 8150. In this same time interval He~II rises and falls similarly to V while the Balmer and He~I rise and flatten off. Near 8020-8030 the He~II $\lambda 4686$, He~I $\lambda 5876$ and Balmer line fluxes seem low compared with the relatively high V-band flux. This suggests that long-term trends may differ between lines and continuum, thus potentially affecting lag measurements based on the faster variations.
\begin{table*}
\centering
\caption{Emission-line widths in the Mean and RMS spectra.}
\label{line_widths}
\begin{threeparttable}
\begin{tabular}{cccccc}
\hline \hline
Line & FWHM(mean) & FWHM(rms) & $\sigma_{\text{line}}$(mean) & $\sigma_{\text{line}}$(rms) & Window\\
     &  (km~s$^{-1}$) & (km~s$^{-1}$) & (km~s$^{-1}$) & (km~s$^{-1}$) & (\AA)\\
\hline
H$\gamma$ & $4402\pm 5$ & $4469\pm 14$ & $1703\pm 1$ & $1786\pm 3$ & $4270-4415$\\\\
He II $\lambda 4686$ & $5271\pm 20$ & $6592\pm 26$ & $2007\pm 4$ & $2282\pm 2$ & $4609-4766$\\\\
H$\beta$ blue side & $4754\pm 4$ & $3273\pm 8$ & $1923\pm 1$ & $1657\pm 3$ & $4782-4944$\\\\
H$\beta$ red side & $4412\pm 5$ & $3874\pm 12$ & $2002\pm 1$ & $1810\pm 3$ & $4782-4944$\\\\
He I $\lambda 5876$ & $3948\pm 11$ & $4408\pm 30$ & $2044\pm 4$ & $2037\pm 12$ & $5751-6006$\\
\hline
\end{tabular}
\begin{tablenotes}
\item[]
\end{tablenotes}
\end{threeparttable}
\end{table*}
\section{Time-series Analysis}
\label{time_series_analysis}
\subsection{Variability Amplitudes}
We characterize the variability of each light curve by measuring the fractional rms amplitude, $F_{\rm var}$, defined as
\begin{equation}
    F_{\rm var} = \dfrac{\sqrt{\sigma_{\ell}^2 - \delta_{\rm rms}^2}}{\left<\ell\right>}\,,
\end{equation}
where $\left<\ell\right>$ and $\sigma^2_{\ell}$ are the mean and variance of the light curve, the $\delta_{\rm rms}^2$ is the rms uncertainty on the measured fluxes \citep{1997ApJS..110....9R, 2002ApJ...568..610E}. The calculated $F_{\rm var}$ and its uncertainty for each broad emission-line light curve are tabulated in Table \ref{ccf_broad_lines_results}. The variability amplitude of about 10\% is considered adequate for reverberation studies \citep{1997ApJS..110....9R}. He~II $\lambda 4686$ exhibits substantial ($30\%$~rms) variability. Correlated variations are evident in the Balmer and He~I
lightcurves despite their lower fractional variations ($\sim8-11\%$). We also calculate $F_{\rm var}$ for the continuum light curves (Figure \ref{cont_lcurves}) and list the results in Table \ref{ccf_continuum_results}. We generally find $F_{\rm var}$ values that decrease with increasing wavelength likely due to the variable component of the light from the accretion disc being bluer than the non-variable components, such as host galaxy starlight with the exception of $i'$ band.  
\subsection{Cross-Correlation Analysis}
\label{iccf}
\begin{figure*}
\centering
\includegraphics[width = \textwidth]{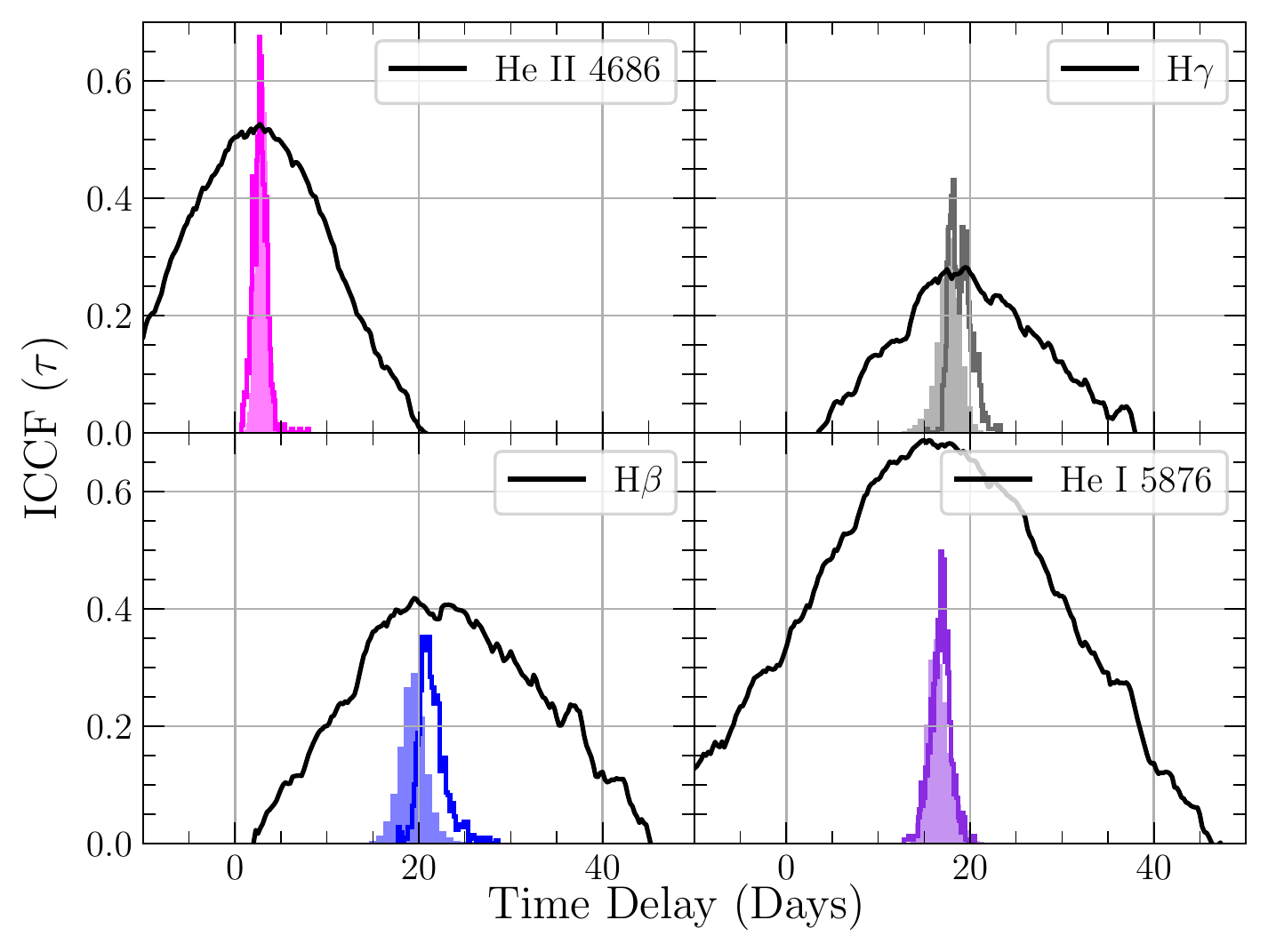}
\caption{The solid lines are the ICCFs $r(\tau)$, and corresponding cross-correlation centroid distributions (CCCD) defined by the FR/RSS samples (solid lines histograms). The translucent histograms are the posterior delay distributions (also shown in Figure \ref{tophat_response_lightcurves} panels a-e, defined by CREAM's MCMC samples of the mean delay of top-hat response functions. These results are for FTN spectra only.}
\label{ccfs}
\end{figure*}

To estimate the time delay $\tau$ of the broad emission-lines (H$\gamma$, He~II, H$\beta$, He~I) relative to the continuum light curve, we used the interpolation cross-correlation function \citep[ICCF,][]{1986ApJ...305..175G, 1987ApJS...65....1G, 1994PASP..106..879W} method. We cross-correlated the emission-line light curves (Figure~\ref{line_lcurves}) with the V-band continuum light curve (Figure~\ref{cont_lcurves}). We chose the V-band continuum light curve because it has more observations and better signal-to-noise than the other continuum light curves in Figure \ref{cont_lcurves}.\\
\indent For two light curves $x(t)$ and $y(t)$ sampled at times $t_i$, the cross-correlation function ICCF~$(\tau)$ is given by
\begin{equation}
\text{ICCF}\,(\tau) = \dfrac{1}{N}\frac{\sum_{i}\left(x(t_{i}) - \bar{x}\right)\left(y(t_{i} - \tau)- \bar{y}\right)}{\sigma_{y}\sigma_{x}}\,,
\end{equation} 
where $\bar{x}$ and $\sigma_x$ are the mean and standard deviation of the $N$ lightcurve samples $x(t_i)$, and similarly for $y$. In the ICCF, the cross-correlation coefficient is calculated twice. Firstly, by cross-correlating the original continuum light curve $y(t_i)$ with the linearly interpolated emission-line light curve $x(t_{i} - \tau)$ for a given lag $\tau$. Secondly, by cross-correlating the linearly interpolated continuum light curve $y(t_{i} - \tau)$ with the original emission-line light curve $x(t_i)$ for a given lag. The ICCF~$(\tau)$ is then taken to be the mean of the two results. We explored a time lag range from $-10$ to 50 days using the interpolation grid spacing of 0.25 days. We based our measured lags on the ICCF centroid $(\tau_{\text{cen}})$ calculated using all $r$ values above a threshold $r > 0.8r_{\text{max}}$ \citep{2004ApJ...613..682P}, where $r_{\text{max}}$ is the maximum value of the correlation coefficient.\\
\indent Uncertainties on the measured lags were determined using the flux randomization and random subset selection method \citep[FR/RSS,][]{1998PASP..110..660P} in the form of PyCCF\footnote{https://bitbucket.org/cgrier/python\textunderscore ccf\textunderscore code}\citep{2018ascl.soft05032S}. For each realization or iteration, individual points are sampled randomly from the original set with replacement, thus resulting in some of the points being excluded and others being selected more than once. In the resulting set, the duplicate points are ignored, which then reduces the set to typically $\sim 37\%$ of the original set. The remaining points are then perturbed by adding random Gaussian deviates based on the associated error and each time calculating $\tau_{\text{cen}}$. This procedure was repeated $10^{3}$ times to build up a cross-correlation centroid distribution (CCCD, Figure \ref{ccfs}) which yielded the median lag and the $68\%$ confidence interval.\\
\indent In our ICCF analysis, we omit the FTS data. When using both FTS and FTN spectra, which have slightly different wavelength scales and flux calibrations, the resulting merged emission-line lightcurves had small but significant jumps
between the FTN and FTS data points. This adversely affected the
ICCF analysis. Therefore we opted to use just the FTN spectra,
which had better time sampling than FTS. We summarise the resulting ICCF lag measurements in Table \ref{ccf_broad_lines_results} (and Figure \ref{ccfs}).
\begin{table}
\caption{3C~120 broad emission-line lags relative to V.}
\label{ccf_broad_lines_results}
\begin{threeparttable}
\begin{tabular}{ccccc}
\hline \hline
Line\\Curve & $r_{\rm max}$ & $\tau_{\text{cent}}$ & $\tau_{\text{CREAM}} - \tau_{\text{V}}$ & $F_{\text{var}}$\\
     & & (days) & (days) &\\
\hline
H$\gamma$ & $0.28$ & $18.8_{-1.0}^{+1.3}$ & $17.6\pm 1.3$ & $0.110\pm 0.008$\\\\
He~II & $0.53$ & $2.7_{-0.8}^{+0.7}$ & $3.0\pm 0.7$ & $0.296\pm 0.022$\\\\
H$\beta$ blue-side & $0.42$ & $21.2_{-1.0}^{+1.6}$ & $19.4\pm 1.7$ & $0.083\pm 0.006$\\\\
H$\beta$ red-side & $0.38$ & $21.2_{-1.0}^{+1.6}$ & $20.2\pm 2.4$ & $0.089\pm 0.006$\\\\
He~I & $0.69$ & $16.9_{-1.1}^{+0.9}$ & $16.6\pm 1.3$ & $0.102\pm 0.007$\\
\hline
\end{tabular}
\begin{tablenotes}
\item[]$\tau_{\text{V}} = 2.26\pm 0.18$ days
\end{tablenotes}
\end{threeparttable}
\end{table}
The FTN blue and red spectra give consistent H$\beta$ lag measurements with $21.2_{-1.0}^{+1.6}$ and $21.2_{-1.0}^{+1.6}$ days respectively, indicating that our emission-line flux measurements are reliable. The H$\gamma$ lag ($18.8_{-1.0}^{+1.3}$) is consistent with the H$\beta$ measurements. The He lines, He~II $\lambda 4686$ lag ($2.7_{-0.8}^{+0.7}$) and He~I $\lambda 5876$ lag ($16.9_{-1.1}^{+0.9}$) have shorter delays than the Balmer lines. Generally, we measure $r_{\rm max}$ values that are lower for each of the broad lines despite the ICCF being well-behaved. 3C~120 being a radio-galaxy (also a miniblazar), there may be an additional contribution by the optical synchrotron continuum emission from the jet which could affect the correlation coefficient values \citep{2019arXiv190904511L, 2019ApJ...876...49Z}. However, synchrotron emission would introduce features in the continuum light curve, and the low correlation coefficients we see stem from relatively large scatter in the emission-line (particularly H$\gamma$ and H$\beta$) data that are not present in the V continuum light curve. We believe this scatter arises from limitations in the accuracy of our spectrophotometric calibration, rather than to variable jet emission affecting the light curves. To try to improve the coefficients, we applied detrending \citep{1999PASP..111.1347W} prior to ICCF analysis. We then omitted the pre-gap spectra with/without detrending. In all cases, the ICCF results did not improve considerably.\\
\indent In addition to the broad emission-line lags with ICCF, we measure the lag between the continuum bands with respect to V in a similar manner as above (Table \ref{ccf_continuum_results}). We give more emphasis on the outcome and implications in terms of the standard disc model in Section \ref{continuum_lags_accretion_disc}.
\subsection{CREAM Modelling}
\label{CREAM}
We undertook a separate analysis of the light curves in addition to the ICCF, using the Continuum REprocessed AGN Markov Chain Monte Carlo code \citep[CREAM,][]{2016MNRAS.456.1960S}\footnote{Python version from https://github.com/dstarkey23/pycecream} with the aim of comparing the results with the ICCF. The ICCF method works best with regularly-sampled data and accurate uncertainty estimates, while CREAM fits a parameterised model to the lightcurve data, including parameters for the reverberating disc model and a noise model, with 1 or 2 parameters per lightcurve that adjusts the nominal error bars. The noise model optimizes an additive extra variance parameter ($\sigma_0^2$) and multiplicative error-bar scaling factor $(e)$ such that
\begin{equation}
\sigma_{\text{new}} = \sqrt{(e\sigma_{\text{old}})^{2} + \sigma_0^2}\,.
\label{errorbar_rescaling}
\end{equation} 
For the CREAM analysis, we model both FTN and FTS data simultaneously. CREAM fits a lamppost model with driving light curve $X(t)$ to the input continuum light curves $f_{\nu}(\lambda, t)$ with the continuum response function $\psi(\tau|\lambda)$,
\begin{equation}
f_{\nu}(\lambda, t) = \bar{F}_\nu(\lambda) + \Delta F_{\nu}(\lambda)\int X\,(t - \tau)\,\psi(\tau|\lambda)d\tau\,,
\label{disc_flux}
\end{equation}
in order to infer the shape of the true driving light curve that drives the continuum variability. The driving lightcurve $X(t)$ is normalised to $\left<X\right>=0$ and $\left<X^2\right>=1$, so that $\bar{F}_\nu(\lambda)$ and $\Delta F_\nu(\lambda)$ represent the mean and rms
spectra respectively. The continuum response function is normalised to $\int \psi(\tau|\lambda)\, d\tau = 1$. For the continuum delay distribution, we adopt a face-on (inclination $i = 0$) steady-state blackbody accretion disc with a power-law temperature profile
\begin{equation}
	T=T_1\,\left( \frac{ r } { r_1 } \right)^{-\alpha}\ ,
\end{equation}
where $T_1$ is the temperature at $r_1=1$~light day,
and $T\propto r^{-\alpha}$ corresponds to $\tau\propto\lambda^{1/\alpha}$.
For the steady disc model, $\alpha=3/4$ and $\tau\propto\lambda^{4/3}$ 
and 
\begin{equation}
	T_1^4 = \frac{ 3\,G\,M\,\dot{M}}{8\,\pi\,\sigma\,r_1^3 }
\ ,
\label{temperature_equation}
\end{equation}
for a given black hole mass $M$ and mass accretion rate $\dot{M}$.
Thus $\tau(\lambda)$ tests the prediction $\alpha=3/4$
and $T_1$ then measures the product $M\,\dot{M}$.\\
\indent CREAM assumes the origin of accretion disc variability is a centrally located point source with a stochastic luminosity modelled as a time series $X(t)$ given by
\begin{equation}
\label{eqdrive}
X \left( t \right) = \sum\limits_{k=1}^{N_k}C_k \cos(\omega_k t) + S_k \sin(\omega_k t),
\end{equation}
where $\omega_k\equiv k\,\Delta\omega$ are equally-spaced
fourier frequencies, with $C_k$ and $S_k$ the corresponding
fourier amplitudes that describe the shape of the lamppost light curve variations. To incorporate prior knowledge about the character of AGN lighcurves, CREAM includes priors on the $S_k$ and $C_k$ parameters that steer the power spectrum of the driving light curve toward a random walk, with fourier power-density spectrum $P(\omega) \propto \omega^{-2}$. These random walk priors are gaussian with mean $\left<S_k\right> = \left<C_k\right> = 0$ and standard deviation $\sigma_k$, where
 \begin{equation}
\label{eqpriorsimp}
\sigma_k^2 = \langle S_k^2 \rangle + \langle C_k^2 \rangle = P(\omega_k)\,\Delta \omega=P_0\,\Delta \omega\,\left( \frac{ \omega_0 }{\omega_k} \right) ^{2}\,.
\end{equation}
CREAM's fitted parameters and their priors are summarized in Table \ref{tabprior}.
\begin{table}
\center
\caption{Summary of priors on each of the CREAM parameters.}
\label{tabprior}
\begin{threeparttable}
\begin{tabular}{ccc}
\hline
Parameter & $N_\mathrm{par}$ & Prior\\
\hline 
\hline
$S_k$ and $C_k$ & 2$N_k$ & Gaussian ($\langle S_k \rangle = \langle C_k \rangle = 0$,  \\
&& $\langle S_k^2 \rangle = \langle C_k^2 \rangle = \sigma_k^2$)$^{\text{a}}$\\
$i$ & 0 & 0\\
$T_1$ & 1 &Log Uniform \\
$\alpha$ & 0 & 3/4 \\
$\Delta F_\nu(\lambda)$ & $N_{\lambda} \times N_\mathrm{telescopes}$ & Log Uniform \\
$\overline{F}_{\nu}\left( \lambda \right)$ & $N_{\lambda} \times N_\mathrm{telescopes}$ & Log Uniform \\
$\bar{\tau}(\lambda)^{\text{b}}$ & $N_{\lambda} $ & Uniform \\
\hline
\end{tabular}
\begin{tablenotes}
\item[a] $\sigma_k$ is defined in Equation \ref{eqpriorsimp}.
\item[b]The top-hat centroid lag parameter is optional and used as a substitute for CREAM's disc response function parameters $T_1$, $\alpha$, and $i$ if CREAM is run in 'top-hat' mode.
\end{tablenotes}
\end{threeparttable}
\end{table}
\\
\indent Following \cite{2017ApJ...851...21G} we use
CREAM to fit multiple continuum lightcurves simultaneously with multiple emission-line lightcurves. For each line we adopt a uniform (top hat) delay distribution specified by mean delay $\bar{\tau}$ and delay width $\Delta\tau$. This is based on the assumption that the BLR responds to changes by the driving light curve. CREAM treats the emission-line light curves as delayed and smoothed versions of the inferred driving light curve with a top hat response function similar to that used in JAVELIN \citep{2011ApJ...735...80Z}. CREAM estimates parameter uncertainties using a Markov chain, in our case with $10^5$ samples.\\
\indent In modelling the light curves, we fit only the temperature $T_1$ at 1 light day from the black hole, holding the power-law index $\alpha\equiv 3/4$ and face-on inclination $i=0$. We used our derived black hole mass $M = 6.3\times 10^{7}\,\text{M}_{\odot}$ (Section~\ref{black_hole_mass}) for consistency. We also set the upper frequency limit to 0.5 cycles per day for the driving light curve, the width of the top hat response to 2 days and restricted the time delay to be within -10 to 50 days. The CREAM fits to the continuum and line light curves are shown in Figures \ref{cream_continuum_lightcurves}-\ref{tophat_response_lightcurves}, along with the inferred driving light curve. The model yielded a fit that is a reasonable representation of most of the observed features in the light curves with a few exceptions here and there. The U-band light curve, Figure \ref{cream_continuum_lightcurves} (panel~h), exhibits large error bars due to atmospheric extinction.\\
\indent The measured lags and their uncertainties are shown in Table \ref{ccf_broad_lines_results} and Table \ref{ccf_continuum_results}, relative to V-band in the observed frame. Since we are measuring relative to V-band, we therefore need to subtract this continuum lag from other CREAM values in order to compare CREAM lags to those obtained with the ICCF method. We find that the two methods give consistent results to within uncertainties for H$\gamma$, He II $\lambda 4686$, He~I $\lambda 5876$ and H$\beta$ lags. We interpret the continuum lags obtained with both ICCF and CREAM in terms of the accretion thin-disc model in Section \ref{continuum_lags_accretion_disc}. 
\begin{table}
\caption{3C~120 continuum lags relative to V in the observed frame.}
\label{ccf_continuum_results}
\begin{threeparttable}
\begin{tabular}{ccccc}
\hline \hline
Line\\Curve & $\lambda$ & $\tau_{\text{cent}}$ & $\tau_{\text{CREAM}} - \tau_{\text{V}}$ & $F_{\text{var}}$\\
     & (\AA) & (days) & (days) &\\
\hline
U & 3656 & $0.14_{-0.36}^{+0.35}$ & $-0.94\pm 0.21$ & $0.114\pm 0.005$\\\\
$g'$ & 4770 & $-0.63_{-0.13}^{+0.13}$ & $-0.38\pm 0.24$ & $0.092\pm 0.003$\\\\
V & 5477 & $0.00_{-0.12}^{+0.13}$ & $0.00\pm 0.26$ & $0.080\pm 0.002$\\\\
$r'$ & 6231 & $1.50_{-0.24}^{+0.24}$ & $0.42\pm 0.28$ & $0.060\pm 0.002$\\\\
$i'$ & 7625 & $2.60_{-0.24}^{+0.24}$ & $1.25\pm 0.34$ & $0.070\pm 0.002$\\\\
$z_s$ & 8660 & $2.86_{-0.24}^{+0.24}$ & $1.9\pm 0.4$ & $0.059\pm 0.002$\\
\hline
\end{tabular}
\begin{tablenotes}
\item[]$\tau_{\text{V}} = 2.26\pm 0.18$ days
\end{tablenotes}
\end{threeparttable}
\end{table}
\section{Inter-band Continuum Lags}
\label{continuum_lags_accretion_disc}
\begin{figure}
\centering
\includegraphics[width = \columnwidth]{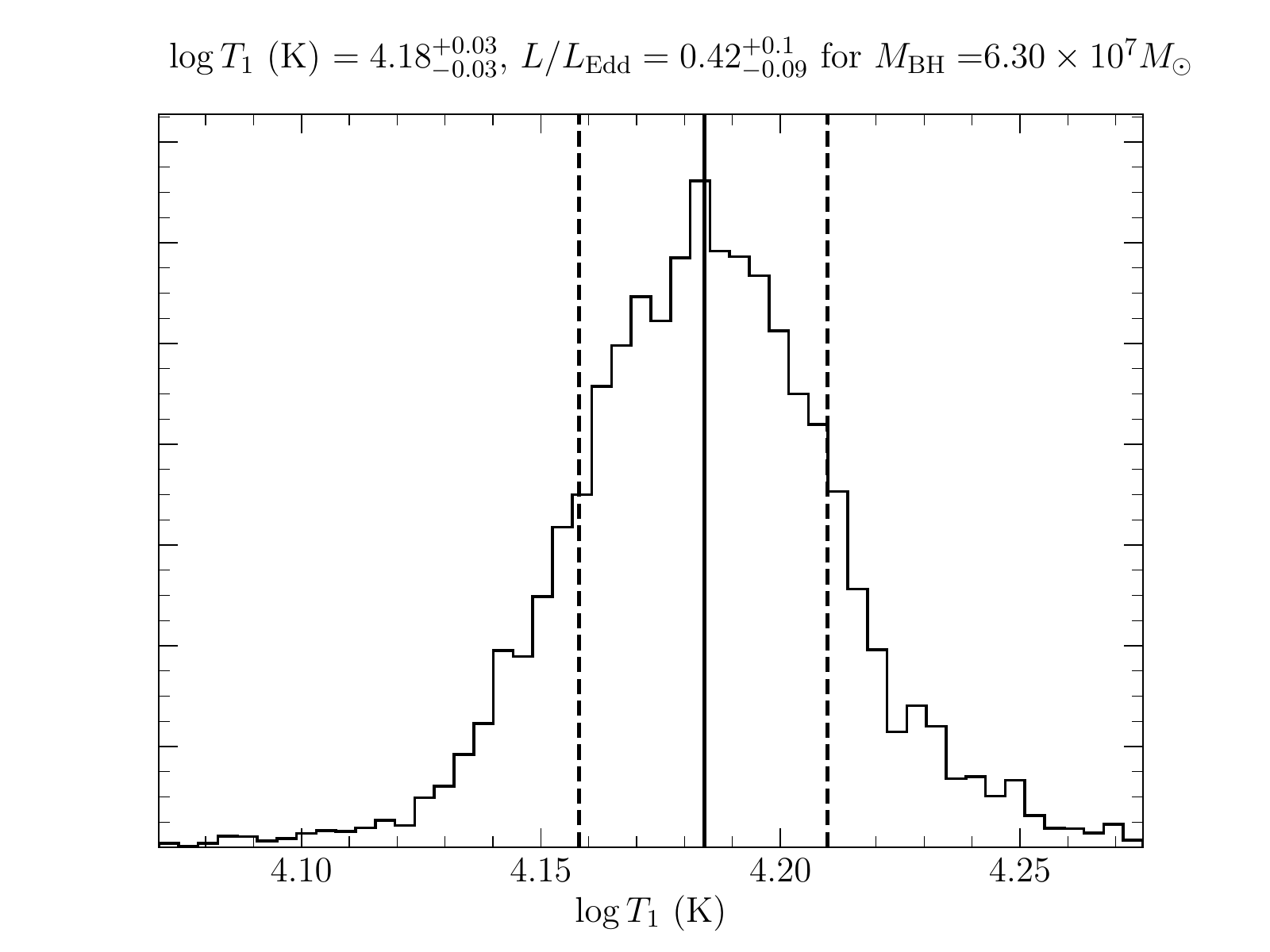}
\caption{The posterior probability histogram for $T_1$ as derived by CREAM, for $\alpha = 3/4$ and $i = 0$.}
\label{temperature_profile}
\end{figure}

We investigate the accretion disc structure in 3C~120 in terms of the standard accretion disc model \citep{1973A&A....24..337S}. In this model, continuum delays increase with increasing wavelength due to thermal reprocessing, where the hotter inner part of the disc responds to the variable ionizing source before the cooler outer part of the disc. The lag is interpreted as the light-travel time from the ionizing source to the reprocessing site, where it is reprocessed into UV/Optical continuum emission. The temperature profile of the accretion disc changes with radius $r = c\tau$ as $T(r)\propto r^{-3/4}$, for a given black hole mass $M$ and accretion rate $\dot{M}$. The wavelength-dependent lags have been detected before in several studies across a wide range of continuum bands \citep[e.g.][]{2007MNRAS.380..669C, 2015ApJ...806..129E, 2016ApJ...821...56F, 2018ApJ...854..107F, 2017ApJ...836..186J} with statistically significant lag detections, as predicted by the reprocessing model. We use CREAM as described in Section~\ref{CREAM} to fit a steady-state blackbody disc model to the U, $g'$, V, $r'$, $i'$, $z_s$ continuum light curves, in order to assess the temperature profile of the disc $T(r)$ in 3C~120, and determine the accretion rate $\dot{M}$ for $M = 6.3\times 10^{7}\,\text{M}_{\odot}$. The resulting CREAM model best-fit parameters are presented in Table~\ref{CREAM_Model_Parameters} and \ref{CREAM_disc_spectrum}. The probability distribution for $T_1$ is shown in Figure \ref{temperature_profile}, yielding the best fit value of $T_1 = (1.51\pm 0.10)\times 10^4$ K. This value of $T_1$ corresponds to $\dot{M} = 0.60\pm 0.15\,M_\odot{\rm yr}^{-1}$ (Eqn.~\ref{temperature_equation}) for our measured BH mass $M=6.3\times10^7$~M$_\odot$, giving $\log{( M\,\dot{M} / {\rm M}_\odot^2{\rm yr}^{-1} )} =7.58\pm0.11$. We also report an Eddington ratio $\dot{m}_{\rm Edd} = \dot{M}/\dot{M}_{\rm Edd} = L/L_{\rm Edd} = 0.42_{-0.09}^{+0.10}$. The ratio $L/L_{\rm Edd}$ is influenced by, (1) the inclination and (2) the adopted BH mass which is uncertain by 0.4~dex.
\begin{table}
\centering
\caption{CREAM model best-fit parameters.}
\label{CREAM_Model_Parameters}
\begin{threeparttable}
\begin{tabular}{cc}
\hline \hline
Parameter & Value \\
\hline
$T_1\,(10^{4}K)$ & $1.51\pm 0.10$\\ 
$i\,(\rm deg)$ & 0 \\
$\alpha$ & $0.75$ \\
\hline
\end{tabular}
\begin{tablenotes}
\item[]
\end{tablenotes}
\end{threeparttable}
\end{table}

\begin{table}
\centering
\caption{Mean and rms Spectra from CREAM.}
\label{CREAM_disc_spectrum}
\begin{threeparttable}
\begin{tabular}{ccc}
\hline \hline
$\lambda$ & $\overline{F}_{\nu}\left( \lambda \right)$ & $\Delta F_\nu(\lambda)$\\
(\AA) & (mJy) & (mJy)\\
\hline
\hline
3656 & $7.604\pm 0.011$ & $0.937\pm 0.013$\\
4770 & $11.308\pm 0.009$ & $0.96\pm 0.01$\\
5477 & $11.80\pm 0.01$ & $0.92\pm 0.01$\\
6231 & $15.516\pm 0.013$ & $0.929\pm 0.013$\\
7625 & $14.317\pm 0.015$ & $0.890\pm 0.013$\\
8660 & $15.442\pm 0.016$ & $0.817\pm 0.015$\\
\hline
\end{tabular}
\begin{tablenotes}
\item[] The dust-extinction-corrected values.
\end{tablenotes}
\end{threeparttable}
\end{table}
\subsection{Lag spectrum}
 
\begin{figure}
\centering
\includegraphics[width = \columnwidth]{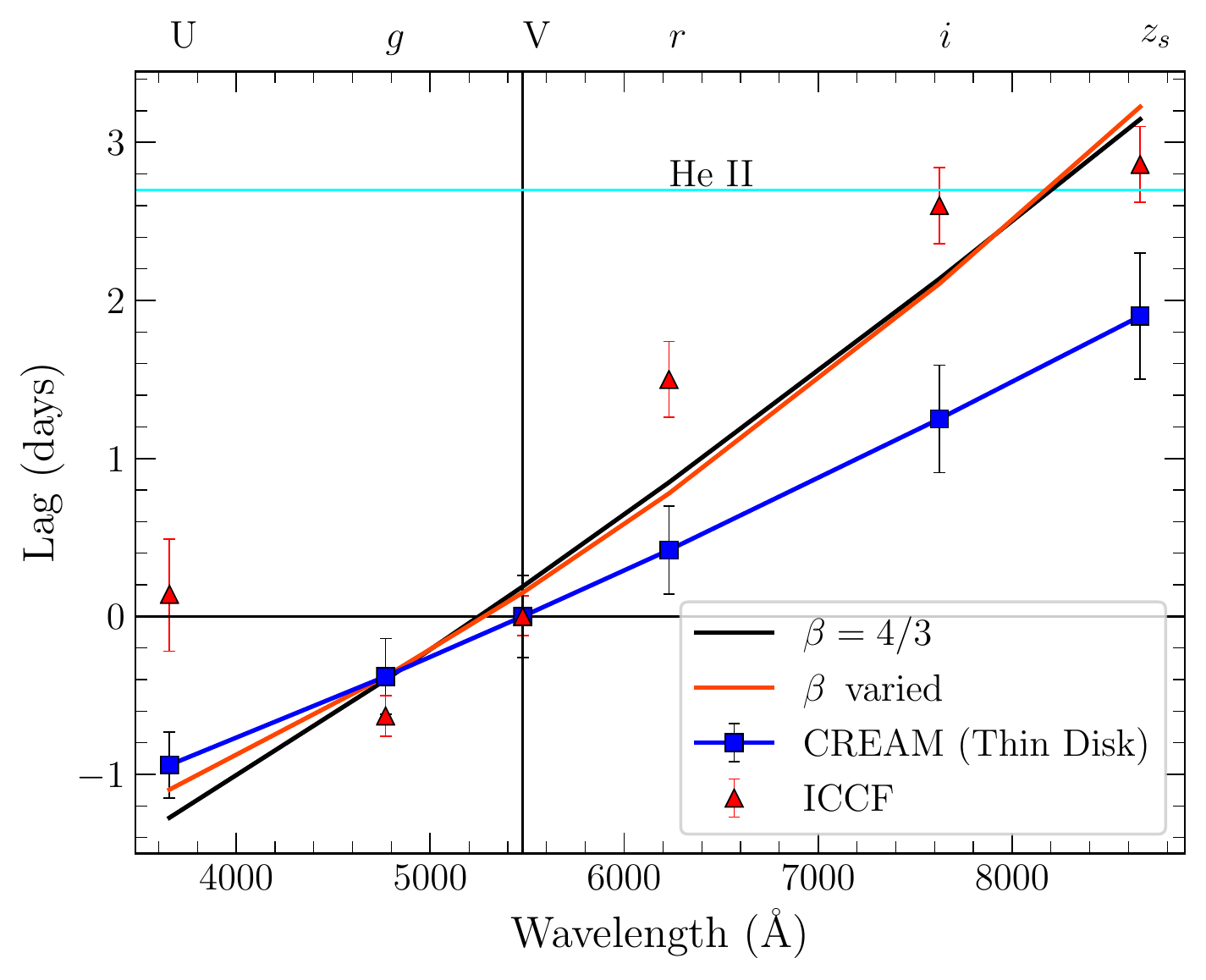}
\caption{Lag as a function of wavelength relative to V band (shown by the vertical line). The best fits of $\tau_0$ with $\beta = 4/3$ (black) and $\beta$ allowed to vary (orange) are shown. The solid blue line is prediction from CREAM standard thin-disc theory with $L/L_{\rm Edd} = 0.4$. The cyan solid line shows the broad He~II emission-line lag of 2.7 days relative to V for comparison.}
\label{lag_versus_wavelength}
\end{figure}

We present the CREAM mean delays alongside ICCF measurements for comparison in Table~\ref{ccf_continuum_results}, and plotted in Figure~\ref{lag_versus_wavelength}. The CREAM lags increase as $\lambda^{4/3}$ since the CREAM model assumes a flat blackbody disc with $T\propto R^{-3/4}$. The ICCF results also show the same trend of rising lag with wavelength, with the exception of the U-band. The U-band resulted in a longer lag than would be expected (i.e. larger than the lag between V and $g'$). This enhanced U lag has been encountered before \citep[e.g.,][]{2015ApJ...806..129E, 2016ApJ...821...56F} and is said to be due to the considerable contribution of Balmer continuum and high-order Balmer lines in the U-band.\\
\indent The observed delay spectrum $\tau(\lambda)$ tests
the accretion disc theory. A thin steady-state blackbody accretion disc has $T^4 = 3\,G\,M\,\dot{M}/8\,\pi\,\sigma\,r^3$ and thus with $T\propto h\,c/k\,\lambda$ for blackbody emission, the predicted delay spectrum should follow
$\tau=r/c \propto \left(M\,\dot{M}\right)^{1/3}\,\lambda^{4/3}$ \citep{1973A&A....24..337S}. More generally, $\tau\propto \lambda^\beta$ corresponds to $T\propto r^{1/\beta}$.

To quantify the lags for a disc reprocessing model, we fit the ICCF  lags with a power law of the form
\begin{equation}
    \tau(\lambda) = \tau_{\rm 0}\left[(\lambda/\lambda_{\rm 0})^{\beta} - 1\right]\,,
\label{power_law}
\end{equation}
where $\lambda_{\rm 0}$ is the reference wavelength, $\beta$ is the power-law index which corresponds to the temperature-radius slope $T\propto r^{-\alpha}$ where $\alpha=1/\beta$ and $\tau_{\rm 0}$ is a normalization factor for measuring the radius of the disc at $\lambda_{\rm 0}$. In Figure~\ref{lag_versus_wavelength}, the orange and black curves show fits of this model
with $\lambda_{\rm 0} = 5477$\text{\AA} and $\beta$ fixed at 4/3 (black) and free to vary (orange). Since there is an error bar on the ICCF V-band lag ($0.00_{-0.12}^{+0.13}$ days), we add a wavelength-independent constant parameter to Eqn. \ref{power_law}. This new parameter allows for non-zero V-band lag, in other words we do not force the fit to have zero lag at V-band wavelength. In doing so, we find best-fit value of $\tau_{\rm 0} = 3.51\pm 0.22$ days from ICCF ($\chi^{2}/{\rm dof} = 8.44$ for dof = 4). When $\beta$ is allowed to vary, we find values of $\tau_{\rm 0} = 2.42\pm1.11$ days and $\beta = 1.79\pm0.62$ from ICCF measurements ($\chi^{2}/{\rm dof} = 10.95$ for dof = 3), see Table~\ref{lag_spectrum_Parameter_fits}. Therefore, we discuss the results for fixed $\beta = 4/3$.  
\begin{table}
\centering
\caption{Lag Spectrum Parameter Fits}
\label{lag_spectrum_Parameter_fits}
\begin{threeparttable}
\begin{tabular}{ccccc}
\hline \hline
Wavelength & $\tau_0$ & $\beta$ & $\chi^{2}/{\rm dof}$ & dof\\
\hline
All & $3.51\pm 0.22$ & 4/3 & 8.44 & 4\\
    & $2.42\pm1.11$ & $1.79\pm0.62$ & 10.95 & 3\\   
\hline
\end{tabular}
\begin{tablenotes}
\item[] 
\end{tablenotes}
\end{threeparttable}
\end{table}

\indent We find that the function gives a poor fit to the ICCF lags owing to the excessive U-band lag and potentially excessive $r'$ and $i'$ lags as well. The large ICCF lags are evident from Figure~\ref{lag_versus_wavelength}, with lags of $i'$, $z_s$ being comparable to and larger than that of the high-ionization-state lines such as He~II $\lambda 4686$. Comparing the intercepts to infer the radius of the disc at $\lambda_{\rm 0}$, $\tau_{\rm 0} = 3.51\pm 0.22$ days for ICCF lags versus $\tau_{\rm 0} = 2.26\pm 0.18$ days for CREAM, implies a radius that is a factor of $1.55\pm 0.16$ times larger than would be expected from the standard thin-disc model. This discrepancy could be due to several different sources: (1) difficulties in the ICCF in measuring the lags due to slightly noisier U light curve (SNR = 50), influence by outliers, or over influenced by the observing gap, (2) contamination of the disc continuum emission by HI bound-free continuum emission from the BLR \citep{2001ApJ...553..695K, 2018MNRAS.481..533L, 2019NatAs...3..251C, 2019MNRAS.489.5284K}. This produces a composite delay distribution, with a sharp peak at small lags from the disc continuum, and a broader peak with longer delays from the BLR. Evidence for this HI bound-free continuum from BLR is shown by the Balmer jump at 3640\AA~and Paschen jump at 8200\AA~ in the ICCF lag spectrum. We do not account for diffuse continuum contribution to the measured continuum lags, we leave that here for future work.  

\begin{figure*}
\centering
\includegraphics[width = \textwidth]{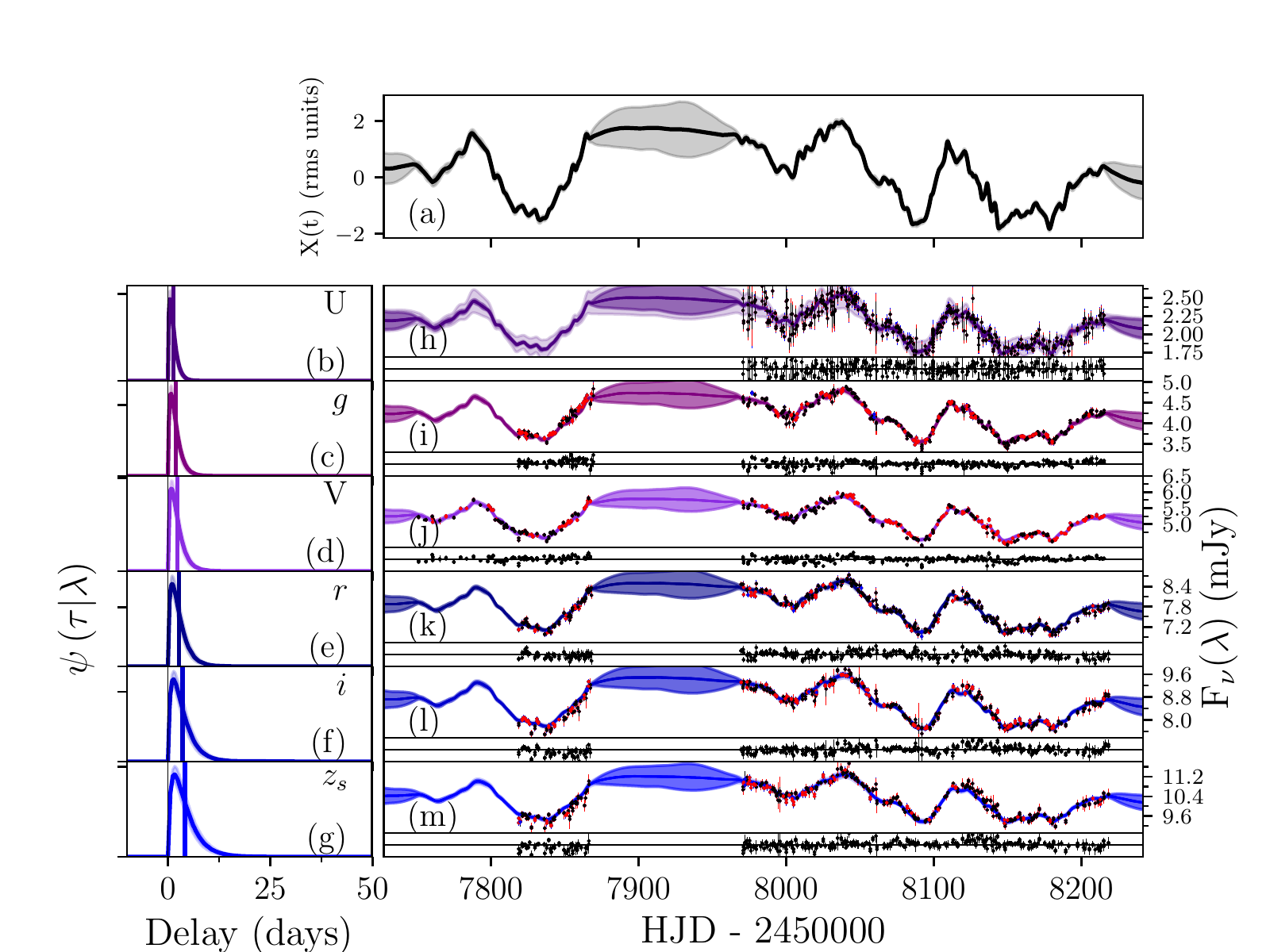}
\caption{CREAM fits to the continuum light curves (panels h-m) from 3C~120. The upper figure (panel~a) shows the inferred driving light curve. Panels b-g show the response functions where the vertical line denote the mean lags. Also, shown are the fit residuals in panels h-m. The red and blue are the rescaled error bars using Eqn.~\ref{errorbar_rescaling}.}
\label{cream_continuum_lightcurves}
\end{figure*}
\begin{figure*}
\centering
\includegraphics[scale = 1.0]{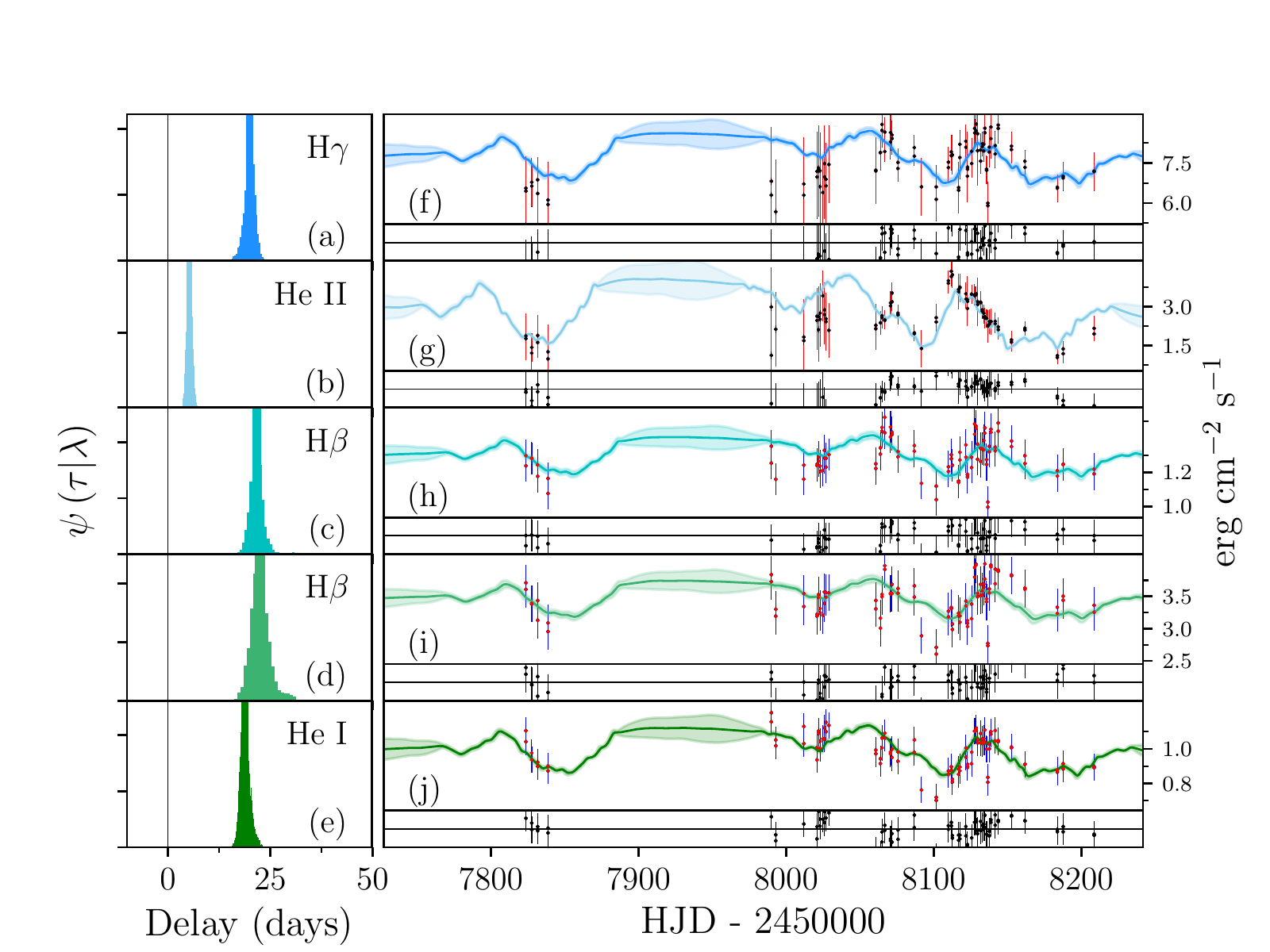}
\caption{The inferred driving light curve, panel~a in Figure~\ref{cream_continuum_lightcurves} is also fit to the line light curves (panels f-j). Panels a-e show the posterior delay distributions, defined by CREAM's MCMC samples of the mean delay of top-hat response functions. The vertical line marks the zero delay. The light curves are H$\gamma\,\lambda 4340$, He II $\lambda 4686$, H$\beta\,\lambda 4861$ (blue-side), H$\beta\,\lambda 4861$ (red-side) and He I $\lambda 5876$ respectively. The units are $10^{-13}$ erg~cm$^{-2}$~s$^{-1}$ for H$\gamma$ and He~II, and $10^{-12}$ erg~cm$^{-2}$~s$^{-1}$ for the other panels.}
\label{tophat_response_lightcurves}
\end{figure*}
\begin{figure}
\centering
\includegraphics[width = \columnwidth]{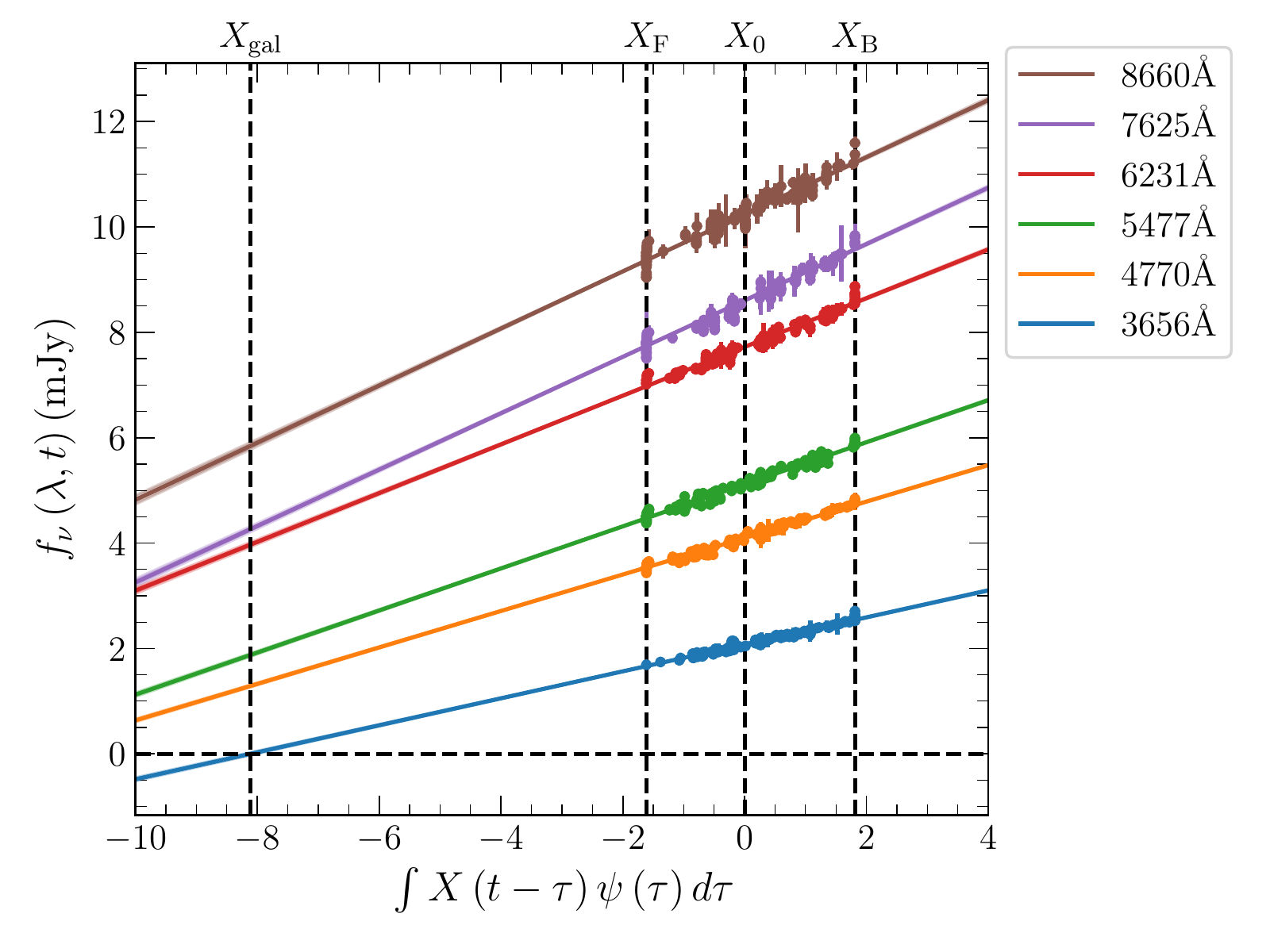}
\caption{Model response light curves for the CREAM fits to the input light curves (not dust-corrected). The vertical lines label the driving light curve values used to evaluate the galaxy, faint-state, and bright state spectra $(X_{\rm gal}, X_{\rm F}, X_{\rm B})$. At $X(t) = X_{\rm gal} = -8.12$ the linear fit crosses zero at the shortest wavelength, placing a lower limit on the host-galaxy flux.}
\label{flux-flux}
\end{figure}
\subsubsection{Accretion-disc Spectrum}
\begin{figure}
\centering
\includegraphics[width = \columnwidth]{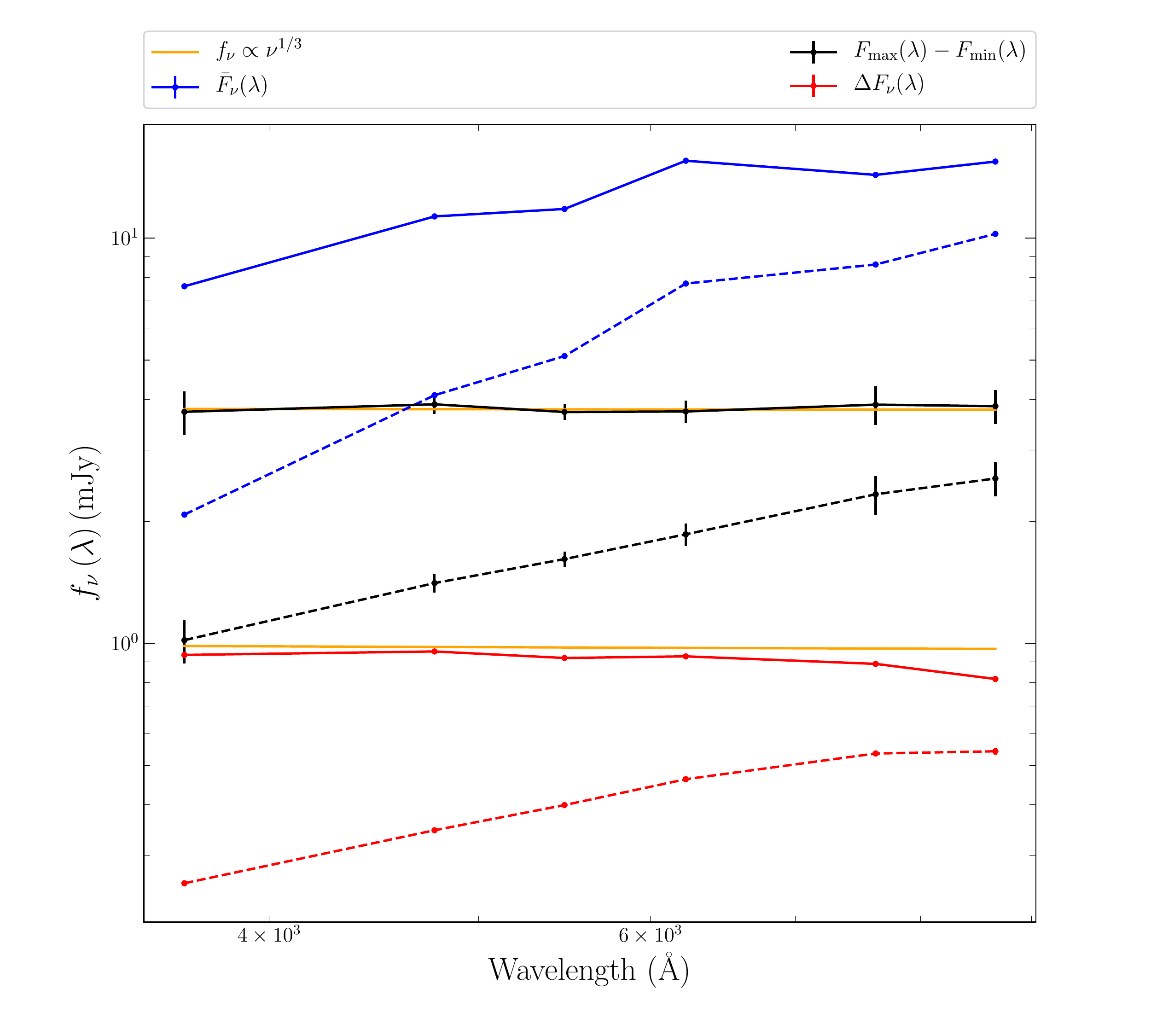}
\caption{The log-log plot of the mean (disc+galaxy) spectrum (blue), rms disc spectrum (red) and max-min (black) disc spectrum, compared with the expected power-law slope (yellow). The dashed lines correspond to reddened spectra and solid lines to de-reddened spectra.}
\label{log-log}
\end{figure}

\begin{figure}
\centering
\includegraphics[width = \columnwidth]{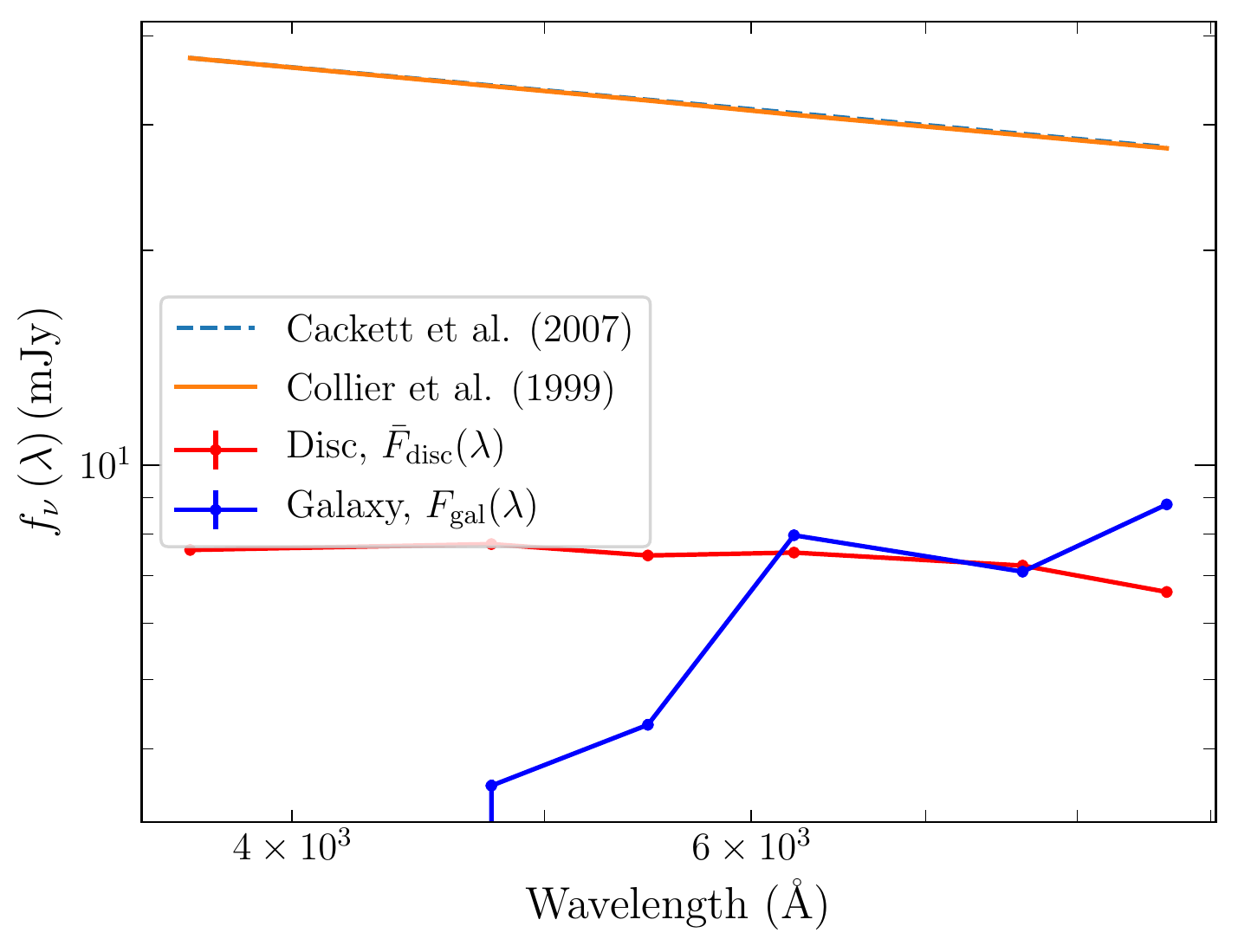}
\caption{The predicted spectra are shown in blue (dotted) and orange (solid) estimated from the $T(r)$ profile derived from the CREAM fit to $\tau(\lambda)$, compared with the mean disc spectrum shown in red. The host-galaxy spectrum is in blue (solid).}
\label{predicted_spectra}
\end{figure}

The linearised reprocessing model assumes that continuum variations respond linearly to the lamppost with a time delay that depends on light-travel time effects. We see from the CREAM fit shown in Figure~\ref{cream_continuum_lightcurves} that an approximate linear relation exists between the inferred driver and disc light curves (both experience correlated variations). We explore this further in Figure~\ref{flux-flux}. The response light curves are plotted relative to the driving light curve $X(t-\tau)$, averaged over the time delay distribution $\Psi(\tau)$ to demonstrate this linear relation. Figure~\ref{flux-flux} shows that this relation holds across all continuum bands indicating that, despite larger than expected delays, a linearised echo description is apparent from correlated variability of these light curves.\\
\indent In Figure~\ref{log-log}, we plot the rms disc spectrum (variable component of the light) derived from the CREAM fit slopes in Figure~\ref{flux-flux} and denoted as $\Delta F_\nu(\lambda)$ in Eqn.~\ref{disc_flux} and Table~\ref{CREAM_disc_spectrum}. The average (disc+galaxy) spectrum is derived from the CREAM fit intercepts in Figure~\ref{flux-flux} and denoted as $\overline{F}_{\nu}\left( \lambda \right)$. Also, shown is the $F_{\rm max}(\lambda) - F_{\rm min}(\lambda)$ spectrum corresponding to the difference between maximum and minimum observed fluxes for each wavelength. As a reference, we overplot the fiducial $f_{\nu}\propto\nu^{1/3}$ power-law disc spectrum (yellow). From the observed fluxes (dashed lines), it appears that the rms disc spectrum is red rather than blue. We attribute this to extinction and reddening by dust in the Milky Way (MW). We correct for MW extinction using a dust extinction map \citep{1998ApJ...500..525S} with $E(B-V) = 0.29\,{\rm mag}$ and an extinction curve of \cite{1999PASP..111...63F} with $R_V = 3.1$. After correcting for MW dust extinction (solid lines), we can see that the variable disc spectrum is blue and quite close to the $\nu^{1/3}$ spectrum predicted by the thin-disc theory, while the average (disc+galaxy) spectrum remains red. We note that there is no obvious sign in the dust-corrected $\Delta F_\nu(\lambda)$ or $F_{\rm max}(\lambda) - F_{\rm min}(\lambda)$ disc spectrum of Figure~\ref{log-log} of Balmer or Paschen jumps that might be expected given their signature in delay spectrum $\tau(\lambda)$ of Figure~\ref{lag_versus_wavelength}.\\
\indent We next compare the predicted disc spectrum calculated from the $T(r)$ profile derived from the CREAM fit to $\tau(\lambda)$, with the observed disc spectrum flux. CREAM fit estimates $T(r)$ from $\tau(\lambda)$ (i.e. $M\dot{M}$ from Figure~\ref{temperature_profile}) and independently of that also estimates the disc flux $f_\nu(\lambda)$ at each wavelength. These are the $\Delta F_\nu(\lambda)$ and $\bar{F}_\nu(\lambda)$ in Eqn. \ref{disc_flux}, which are the slopes and intercepts respectively, plotted as a (MW-dust corrected) spectrum in Figure~\ref{log-log}.
We want to ascertain if the $T(r)$ from the $M\dot{M}$ in Figure~\ref{temperature_profile} predicts a disc spectrum $f_\nu(\lambda)$ that is consistent with (or brighter than, or fainter than) the independently measured disc spectrum in Figure \ref{log-log}. We make use of Eqn. 4 in \cite{2007MNRAS.380..669C} and Eqn. 2 in \cite{1999MNRAS.302L..24C} to derive predicted spectra, and compare with the mean disc $\bar{F}_{\rm disc}$ spectrum (red) in Figure~\ref{predicted_spectra}, where
\begin{align}
    \bar{F}_{\rm disc}(\lambda) = \bar{F}(\lambda) - F_{\rm gal}(\lambda),\\
    F_{\rm gal}(\lambda) = \bar{F}(\lambda) + X_{\rm gal}\Delta F(\lambda).
\end{align}
\indent It follows from Figure~\ref{predicted_spectra} that the predicted disc spectrum
(for $i$=0) is brighter by a factor $\sim4$ compared with the $\bar{F}_{\rm disc}$ spectrum. Also, the shape of the $\bar{F}_{\rm disc}$ spectrum is close to $f_\nu \propto \nu^{1/3}$, in agreement with predicted spectrum for a thin steady-state blackbody disc, but the surface brightness of the disc is lower by a factor $\sim4$ compared with a face-on blackbody disc with the $T(r)$ profile derived from the delay spectrum $\tau(\lambda)$. This discrepancy is in the same direction as, and somewhat larger than, the result found by \cite{2017ApJ...835...65S} for the disc in NGC 5548. It suggests that the effect may be more general and not just a peculiarity of NGC 5548.
\subsubsection{Optical Luminosity at 5100\AA}
Using the flux-flux analysis method from CREAM, we determine the optical luminosity at 5100\AA~corresponding to our monitoring campaign. A linear relationship is observed between the CREAM inferred driving light curve (assuming that continuum variations are driven by a central source of irradiation) and all continuum bands. The best-fit mean flux $\bar{F}$ and rms flux $\Delta F$ for each wavelength are listed in Table~\ref{CREAM_disc_spectrum}, after correcting for galactic extinction following \cite{1998ApJ...500..525S}. The total flux in each filter includes contributions from the disc and host galaxy. We quantify the galaxy's contribution by evaluating the linear trend lines at $X_{\rm gal} = -8.12$ for each filter (Figure~\ref{flux-flux}). The value $X_{\rm gal}$ is defined to be the point where the linear fit extrapolated to low flux first crosses zero. This occurs first at $X_{\rm gal} = -8.12$, for the fit to the shortest-wavelength U-band flux. Table~\ref{host_galaxy_subtracted_fluxes} summarizes the results of decomposing total (disc+galaxy) flux into galaxy and disc (total-galaxy) components. We determine the interpolated host-galaxy flux at 5100\AA~of $3.913\pm 0.077$ mJy.\\
\indent \cite{2006ApJ...644..133B} using high resolution HST images, determined the host-galaxy flux at 5100\AA~(5.22 mJy) after correcting for extinction and adding narrow-line contribution \citep{2010ApJ...711..461S}. \cite{2009ApJ...697..160B} re-modeled the host-galaxy flux at 5100\AA~ and found it to be 3.32 mJy. The difference between these two host-galaxy estimates is linked with the difference in the decomposition modeling techniques of the galaxy \citep{2012A&A...545A..84P}. Through interpolating, we determine the disc (total-galaxy) flux at 5100\AA~to be $f_{\rm obs}\left[5100{\rm \AA}(1 + {\rm z})\right] = 7.55\pm 0.08\,{\rm mJy}$. Using the redshift distance of 139 Mpc, we measure the luminosity at the time of our campaign to be $(9.94\pm0.11)\times10^{43}$ erg~s$^{-1}$ (Table~\ref{tabulated_previous_measurements}). 
\begin{table}
\centering
\caption{Host-galaxy subtracted fluxes for each wavelength (dust-corrected).}
\label{host_galaxy_subtracted_fluxes}
\begin{threeparttable}
\begin{tabular}{cccc}
\hline \hline
$\lambda$ & Total & Galaxy & Total - Galaxy\\
\hline
(\AA)   &  (mJy) & (mJy) & (mJy)\\
\hline
3656 & $7.604\pm0.011$ & 0 & $7.604\pm0.011$\\
4770 & $11.308\pm0.009$ & $3.553\pm 0.075$ & $7.75\pm0.08$\\   
5477 & $11.80\pm0.01$ & $4.32\pm0.08$ & $7.47\pm0.08$\\
6231 & $15.516\pm0.013$ & $7.974\pm0.105$ & $7.54\pm0.11$\\
7625 & $14.317\pm0.015$ & $7.089\pm0.105$ & $7.23\pm0.11$\\
8660 & $15.442\pm 0.016$ & $8.806\pm0.119$ & $6.64\pm0.12$\\
\hline
\end{tabular}
\begin{tablenotes}
\item[] 
\end{tablenotes}
\end{threeparttable}
\end{table}
\section{Black Hole Mass}
\label{black_hole_mass}
\begin{figure}
\centering
\includegraphics[width = \columnwidth]{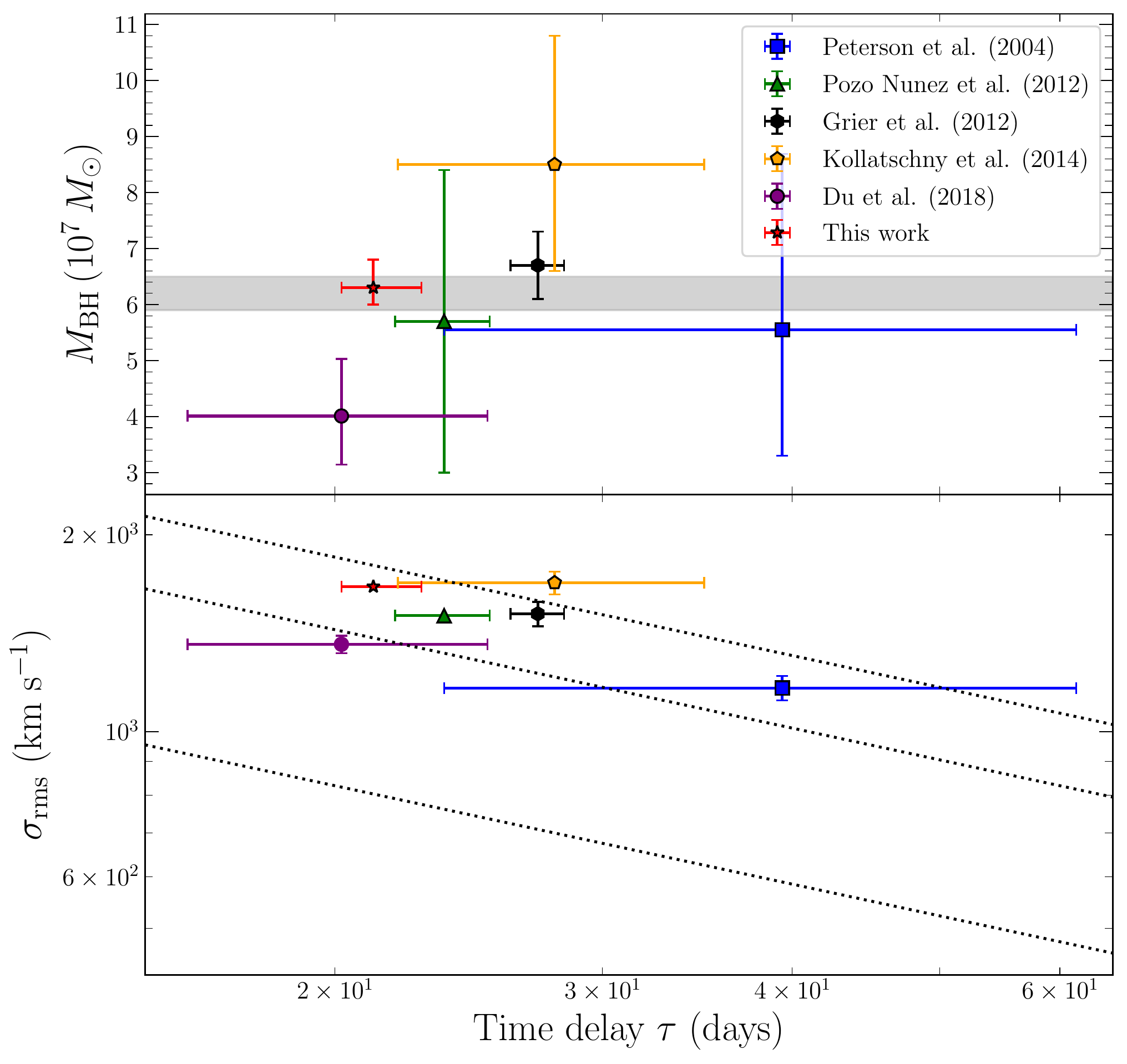}
\caption{Top: The BH mass as a function of the H$\beta$ time delay, shown are the different campaigns on 3C~120. The gray band indicates the $1\sigma$ envelope of the weighted average of the BH mass based on H$\beta$ measurements. Bottom: The H$\beta$ broad emission-line width as a function of the lag. From bottom to top, the dotted lines correspond to constant black hole masses of  $2\times10^{7}{\rm M}_\odot$, $6\times10^{7}{\rm M}_\odot$, $1\times10^{8}{\rm M}_\odot$ respectively for $f = 5.5$.}
\label{compare_BH_masses}
\end{figure}

There have been several previous campaigns \citep[e.g.][]{1998ApJ...501...82P, 2004ApJ...613..682P, 2012A&A...545A..84P, 2014A&A...566A.106K, 2017ApJ...849..146G, 2018ApJ...869..142D} reporting BH mass estimates in 3C~120. To determine the BH mass, we require the emission-line time delay $\tau$ and velocity width $\Delta V$. Combining these two parameters, we can estimate the mass of the BH using Eqn.~(\ref{BH_mass_formula}), thus assuming that the gas motions in the BLR are dominated by the strong gravity of the central BH.\\
\indent We measure the velocity width and its uncertainty from the mean and rms spectra produced by PrepSpec (Section~\ref{prepspec}). Our linewidth measurements are listed in Table~\ref{line_widths}. Our measurements are generally consistent with those measured by \cite{2014A&A...566A.106K} but with better statistical errors. We compare only with \cite{2014A&A...566A.106K} because other papers report linewidths only for H$\beta$.\\
\indent We use the velocity dispersion $\sigma_{\rm line}$ estimated from the rms spectrum in the calculation of the BH mass since it yields less biased black hole mass measurements than FWHM \citep{2011arXiv1109.4181P}. The dimensionless factor $f$ in Eqn.~\ref{BH_mass_formula} is different for each AGN. For comparison with similar studies on 3C~120, we adopt a mean factor of $\langle f \rangle = 5.5$ \citep{2004ApJ...615..645O} estimated by bringing the $M_{\text{BH}}-\sigma_{*}$ relation into agreement with the same relation for quiescent galaxies. We also use our H$\beta$ lag of $21.2_{-1.0}^{+1.6}$ days relative to the V-band continuum and linewidth ($\sigma_{\rm line} = 1657\pm 3$ km~s$^{-1}$), to calculate the black hole mass. We measured our resulting mass of 3C~120 to be $M=\left(6.3^{+0.5}_{-0.3}\right)\times10^7\,(f/5.5)$~M$_\odot$.\\ 
\indent We compare our black hole mass estimate based on H$\beta$ with the individual campaigns where the H$\beta$ line was used to measure black hole mass in 3C~120 (Table~\ref{tabulated_previous_measurements} and Figure~\ref{compare_BH_masses}). \cite{2004ApJ...613..682P} did a reanalysis of 3C~120 data and reported the H$\beta$ lag of $39.4_{-15.8}^{+22.1}$ days with large uncertainty, accounting for the large errors on their derived black hole mass $M = \left(5.55^{+3.14}_{-2.25}\right)\times10^7\,(f/5.5)$~M$_\odot$. \cite{2012ApJ...755...60G} deduced a lag of $27.2_{-1.1}^{+1.1}$ days from JAVELIN, which performs better statistically especially on well-sampled data as in their campaign. Hence, the smaller uncertainties and led to a BH mass $M=\left(6.7^{+0.6}_{-0.6}\right)\times10^7\,(f/5.5)$~M$_\odot$. Recently, \cite{2018ApJ...869..142D} detected a lag of $20.2_{-4.2}^{+5.0}$ days which is similar to our H$\beta$ lag in this work, and measured the black hole mass $M=\left(3.26^{+0.83}_{-0.71}\right)\times10^7\,(f/4.47)$~M$_\odot$. We combine the mass estimates and uncertainties from all measurements to produce the weighted mean \citep{2003sppp.conf..250B} of the black hole mass $M=\left(6.2^{+0.3}_{-0.3}\right)\times10^7\,(f/5.5)$~M$_\odot$. We note that these BH mass measurements incorporate only the measurement uncertainties in $\tau$ and $\sigma_{\text{line}}$, they do not take into account the uncertainty in $f$. We intentionally did not include \cite{1998ApJ...501...82P} as no uncertainties were determined on their black hole mass. Generally, we find consistent mass estimates to within $1\sigma$ error limits except for \cite{2018ApJ...869..142D}, possibly due to the significant difference in the H$\beta$ linewidth and lag they measure compared to other studies since the black hole mass is a combination of these two parameters. We also plot the measured H$\beta$ lags and rms linewidth in Figure~\ref{compare_BH_masses} and see that they are generally consistent with the virial assumption (i.e the lags should decrease as the linewidth increases).\\
\indent \cite{2017ApJ...849..146G} measured the BH mass for 3C~120 to be $M = 6.92_{-1.55}^{+1.38}\times 10^{7}\,M_{\odot}$, through dynamical modeling \citep{2014MNRAS.445.3055P, 2014MNRAS.445.3073P} of the object's BLR. This way of measuring the mass does not assume a particular value of $f$ and relies on direct modeling to measure the BH mass. It also eliminates the uncertainty introduced by assuming a value of $f$. Using mass derived from dynamical modeling, the virial factor $f$ can be inferred for individual objects. \cite{2017ApJ...849..146G} estimate $f$ specific to 3C~120 of $f_{\sigma} = 5.75\pm 2.25$ and $f_{\rm FWHM} = 6.46\pm 2.53$, where $\sigma$ is from the rms spectrum and FWHM is measured from the \cite{2012ApJ...755...60G} mean spectrum. The $f$ adopted here is consistent with $f$ inferred from dynamical modeling, leading to consistent black hole masses to within $1\sigma$.  
\begin{table*}
\caption{Outcomes of previous reverberation mapping studies on 3C~120.}
\label{tabulated_previous_measurements}
\begin{threeparttable}
\begin{tabular}{cccccc}
\hline \hline
Study & Campaign & H$\beta$ lag & $\sigma_{\rm line}$ &
$M_{\text{BH}}^{(\text{b})}$ & $\log{\lambda\,L_\lambda(5100\text{\AA})}$\\
      &          &  (days) & (km~s$^{-1}$) & $( 10^{7}\,\text{M}_{\odot})$ & (erg~s$^{-1}$)\\\\
(1) & $1989 - 1996$ & $39.4_{-15.8}^{+22.1}$ & $1166\pm 50$ & $5.55_{-2.25}^{+3.14}$ & $44.01\pm 0.05$\\\\
(2) & $2008 - 2009$ & $27.9_{-5.9}^{+7.1}$ & $1689\pm 68$ & $8.5_{-1.9}^{+2.3}$ & $44.12\pm 0.07$\\\\
(3) & $2009 - 2010$ & $23.6_{-1.7}^{+1.7}$ & $1504^{\rm (c)}$ & $5.7_{-2.7}^{+2.7}$ & $43.84\pm 0.04$\\\\
(4) & $2010 - 2011$ & $27.2_{-1.1}^{+1.1}$ & $1514\pm 65$ & $6.7_{-0.6}^{+0.6}$ & $43.96\pm 0.06$\\\\
(5) & $2016 - 2017$ & $20.2_{-4.2}^{+5.0}$ & $1360\pm 42$ & $4.01_{-0.87}^{+1.02}$ & $43.99\pm 0.01^{\text{\textcolor{red}{*}}}$\\\\
(6) & $2016 - 2018$ & $21.2_{-1.0}^{+1.6}$ & $1657\pm 3$ & $6.3_{-0.3}^{+0.5}$ & $43.99\pm 0.01$\\      
\hline
Mean$^{(\text{a})}$ &  &  &  & $\mathbf{6.2_{-0.3}^{+0.3}}$ &\\
\hline
\end{tabular}
\begin{tablenotes}
\item[](1) \cite{2004ApJ...613..682P}, (2) \cite{2014A&A...566A.106K}, (3) \cite{2012A&A...545A..84P}, (4) \cite{2012ApJ...755...60G}, (5) \cite{2018ApJ...869..142D}, (6) This work.
\item[a] Weighted average.
\item[b] $M$ masses calculated using $\langle f \rangle = 5.5$ for comparison.
\item[c] Measured from single-epoch spectra.
\item[\textcolor{red}{*}] The authors do not provide the measurement, we put that here because the two campaigns overlap.
\end{tablenotes}
\end{threeparttable}
\end{table*}
\section{Velocity-resolved Lag Measurements}
\label{velocity_resolved_lag_measurements}
\begin{figure}
\centering
\includegraphics[width = \columnwidth]{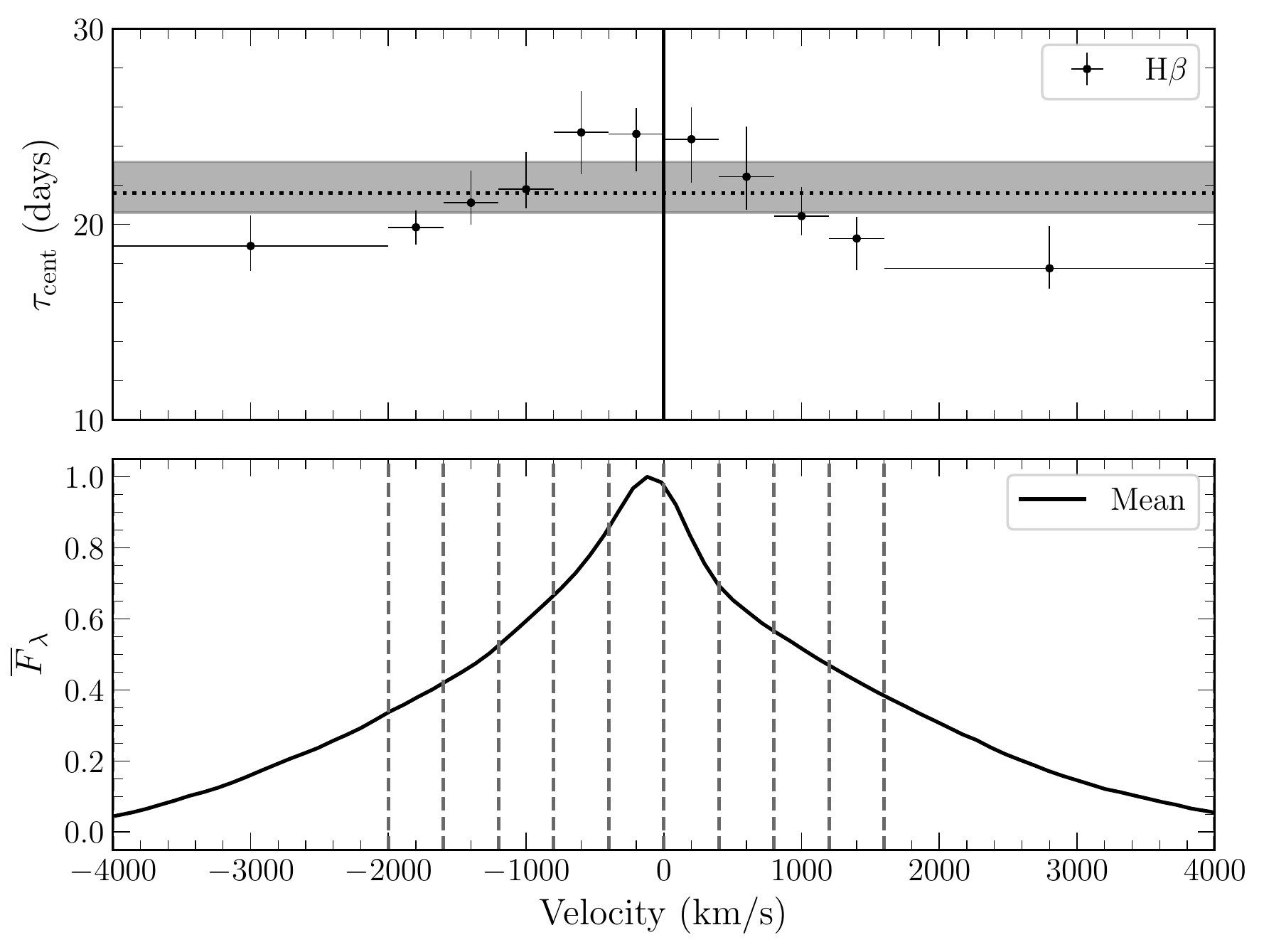}
\caption{Velocity-resolved lags for the broad H$\beta$ emission-line in 3C~120. The upper panel shows the measured centroid lags at different velocity bins. The dotted horizontal line is the average lag and the grey band its uncertainty (from Table \ref{ccf_broad_lines_results}). The lower panel shows the continuum-subtracted normalised mean spectrum. The dotted lines show the bins. Error bars in the horizontal direction represent the velocity bin size.}
\label{velocity_resolved_lag}
\end{figure}

The ultimate goal of RM studies is to ascertain the geometry and kinematics of the BLR through mapping the responsivity of the line-emitting gas as a function of time lag and line-of-sight velocity. Well-conducted velocity-resolved studies can provide information about the distribution and velocity field of the emitting gas within the BLR, whether infall/inflow, outflow and/or Keplerian orbits. In an infall/inflow, longer lags are blueshifted having negative velocities and shorter lags are redshifted having positive velocities. In an outflow, the opposite pattern to that of an infall can be seen. Keplerian/circular orbits are symmetric about zero velocity, with shorter lags on the wings of the profile \citep[e.g.][]{1991ApJ...379..586W, 2009ApJ...705..199B, 2009ApJ...704L..80D, 2010ApJ...721..715D, 2011ApJ...743L...4B, 2013ApJ...764...47G, 2016ApJ...820...27D, 2018ApJ...866..133D}.\\
\indent We undertook velocity-resolved lag measurements across the H$\beta$ emission-line profile in 3C~120. Previous studies \citep{2013ApJ...764...47G, 2017ApJ...849..146G, 2018ApJ...869..142D} have been successful in recovering the structure and kinematics of the BLR in 3C~120 using velocity-resolved lag measurements. We divided the H$\beta$ emission-line profile into 11 velocity bins across the line in the continuum-subtracted spectra from $-4000$ km s$^{-1}$ to $4000$ km s$^{-1}$, using the FTN blue-side data. We set the zero-velocity using the wavelength of the H$\beta$ line in rest frame to separate the blueshifted and redshifted measurements. We constructed H$\beta$ light curves by integrating the line fluxes in each velocity bin.\\
\indent We then measured the time lag $(\tau_{\text{cen}})$ for each of the light curves with respect to the V-band continuum light curve using the ICCF, as in Section~\ref{iccf}. We present our velocity-resolved results in the rest frame in Figure~\ref{velocity_resolved_lag} based on the mean spectrum. Our H$\beta$ velocity-resolved lag profile exhibits a pattern consistent with Keplerian/circular with longer lags at about zero velocity and shorter lags on the wings of the profile. The lag is $\tau\sim25$~days near line centre
and drops symmetrically to $\sim17$~days at $\pm 4000$ km s$^{-1}$. This is consistent with \cite{2013ApJ...764...47G} where they deduced a similar morphology.
\section{discussion}
\label{discussion}
\subsection{Mass Determination}
Because of its nature, 3C~120 has been the subject of 5 previous spectroscopic reverberation mapping studies, thus providing a good sample of independent measurements in the literature to which we compare our results. Based on the ICCF and CREAM methods, we measure the lags of H$\gamma$, He~II $\lambda 4686$, H$\beta$, and He~I $\lambda 5876$ in 3C~120 (Table~\ref{ccf_broad_lines_results}). We find that both methods give consistent lag measurements and small uncertainties. We therefore adopt lags from ICCF as our primary method.\\
\indent We list the variability amplitudes for each of the broad emission lines in Table~\ref{ccf_broad_lines_results}. He~II $\lambda 4686$ has a higher variability (30\%) and shorter time delay (2~days) than do the Balmer and He~I lines, suggesting it originates really close, within a few light days, of the ionising source. This may also be an indication of the BLR stratification in 3C~120 \citep{2003A&A...407..461K}. We compare our measurements with \cite{2014A&A...566A.106K}, in which multiple-line lag measurements were also reported. We note that He~II $\lambda4686$ lag is similar to V-band lag, suggesting the He~II $\lambda4686$ emitting region is close to V emitting region (similar physical size). So we add back the V-band continuum lag to He~II $\lambda4686$ for comparison with other lines, thus we end up with $5.0_{-0.8}^{+0.8}$ days.\\
\indent Our H$\beta$ measurement, combined with the previous studies indicate that the H$\beta$ line time lag is decreasing since the first campaign (Table~\ref{tabulated_previous_measurements}). \cite{2004ApJ...613..682P} measured a long delay, $40\pm20$~days. Campaigns succeeding that measured shorter delays with much improved error estimates as a result of robust time delay measurement techniques such as JAVELIN in \cite{2012ApJ...755...60G} and better sampling. This decreasing trend in H$\beta$ lag might be expected to correspond to a corresponding decrease in optical luminosity. The H$\beta$ reverberations depend on the continuum level during the time of the monitoring campaign, such that longer time lags would correspond to higher continuum states \citep{2006MNRAS.365.1180C}. This expected correlation is not observed (Table~\ref{tabulated_previous_measurements}).\\
\begin{table}
\centering
\caption{Black hole masses for the individual lines.}
\label{black_hole_masses}
\begin{threeparttable}
\begin{tabular}{cccc}
\hline \hline
Line & $\tau$ & $\sigma_{\rm rms}$ & $M$\\
\hline
     &  (days) & (km~s$^{-1}$) & ($M_\odot$)\\
\hline
(1) & (2) & (3) & (4)\\
\hline
H$\gamma$ & $18.8_{-1.0}^{+1.3}$ & $1786\pm3$ & $6.4_{-0.5}^{+0.3}$\\\\
He~II $\lambda4686$ & $5.0_{-0.8}^{+0.8}$ & $2282\pm2$ & $2.8_{-0.5}^{+0.5}$\\\\   
H$\beta$ & $21.2_{-1.0}^{+1.6}$ & $1657\pm3$ & $6.3_{-0.3}^{+0.5}$\\\\
He~I $\lambda5876$ & $16.9_{-1.1}^{+0.9}$ & $2037\pm12$ & $7.5_{-0.5}^{+0.4}$\\\\
\hline
& & & $\times 10^{7}$\\
\hline
\end{tabular}
\begin{tablenotes}
\item[] $\left<f\right>=5.5$
\end{tablenotes}
\end{threeparttable}
\end{table}
\indent Using the measured centroid lags of the Helium and Balmer lines, and their rms linewidths, we estimate black hole mass from each line, as listed in Table~\ref{black_hole_masses}. We generally find that the black hole masses based on the individual lines are consistent to within $1\sigma$ uncertainties except for He~II $\lambda4686$. The black hole mass is the combination of the lag and velocity width. The rms spectrum in Figure \ref{mean_specs} shows that the red wing of He~II $\lambda 4686$ is blended with the H$\beta$ line, and this blending may lead to underestimation of the linewidth, and hence would affect the final black hole mass measurement.   
\subsection{BLR Geometry and Kinematics}
Our high signal-to-noise and high cadence allowed us to carry out the velocity-resolved time lag analysis to probe the geometry and kinematics of the BLR in 3C~120 across the H$\beta$ emission-line profile. Our velocity-resolved profile is shown in Figure~\ref{velocity_resolved_lag}. We see clearly a symmetric pattern about zero velocity consistent with Keplerian motions from $-4000$~km~s$^{-1}$ to $+4000$~km~s$^{-1}$, with higher lags at about zero velocity and shorter lags on the wings. The line wings have almost identical delays with respect to the V-band continuum. This is interesting as it indicates that the velocity field is not dominated by radial inflow or outflow. There is an overall gravitational domination of the BLR gas motions by the black hole, thus validating the assumption (i.e gravity dominated system) that allow us to measure the mass of the central black hole.\\
\indent \cite{2013ApJ...764...47G} also deduced a kinematics consistent with our findings but with an additional component, in which they interpreted their velocity-resolved lag measurement as a combination of circular orbits plus inflowing gas from the campaign $2010-2011$. In an earlier campaign $(2008-2009)$, \cite{2014A&A...566A.106K} deduced a two-component BLR structure similar to that of \cite{2013ApJ...764...47G, 2017ApJ...849..146G}. This is not surprising given the small time gap between these campaigns. The most recent campaign $(2016-2017)$ is by \cite{2018ApJ...869..142D} in which they deduced from their H$\beta$ velocity-resolved profile, a pattern consistent with an outflow from $1500$ km~s$^{-1}$ to $-1500$ km~s$^{-1}$ in their rms spectrum, although with large error bars.
\subsection{CREAM thin disc model}
In addition to the broad emission-line variability studies, we also undertook the inter-band continuum variability studies in 3C~120 following our high signal-to-noise, densely sampled and highly correlated continuum light curves in six bands spanning the wavelength range from 3656-8660\AA. To investigate further this waveband correlation in terms of the standard thin-disc model, we fit the continuum light curves using CREAM's thin steady-state blackbody disc model ($T=T_1(r_1/r)^{3/4}$ with $T_1^4\propto M\, \dot{M}/r_1^3$), in order to measure the disc temperature $T_1$ at $r=1$ light day, and hence $\dot{M}$ for $M = 6.3\times 10^{7}\,\text{M}_{\odot}$. Generally, the CREAM model fits the data quite well even with the sporadic variability in the broad lines. We find values of $T_1 = (1.51\pm 0.10)\times 10^4$ K, mass accretion rate $\dot{M} = 0.60\pm 0.15\,M_\odot{\rm yr}^{-1}$ and corresponding Eddington ratio $\dot{m}_{\rm Edd} = \dot{M}/\dot{M}_{\rm Edd} = L_{\rm bol}/L_{\rm Edd} = 0.42_{-0.09}^{+0.10}$, implying 3C~120 is accreting at about half the Eddington rate. Section~\ref{discussion_lag_spectrum} summarizes the results of the CREAM mean delays when compared with model-independent ICCF delays in terms of the $\tau-\lambda$ relation. Section \ref{discussion_accretion_disc_spectrum} goes over the accretion disc spectrum and implications thereof.  
\subsubsection{Lag spectrum}
\label{discussion_lag_spectrum}
The ICCF lag spectrum shows lag-wavelength dependence relation consistent with reprocessing model $\tau\propto\lambda^{\beta}$, when allowing the normalization factor $\tau_0$ to vary and fixing $\beta$ to 4/3. The excess U lag (i.e. greater than expected from extrapolating the longer-wavelength lags) is observed in Figure~\ref{lag_versus_wavelength}. The larger U lag than would be expected has been reported in previous inter-band continuum lag studies. For example, \cite{2015ApJ...806..129E} and \cite{2016ApJ...821...56F} also detect this excess U lag, and link it to contamination by the broad emission lines and diffuse continuum (DC) from the BLR \citep[e.g.,][]{2001ApJ...553..695K}. This DC contamination is signified by the Balmer jump at 3640\AA~and Paschen jump at 8200\AA~in Figure~\ref{lag_versus_wavelength}. This contamination tends to lead to disc sizes several times larger than expected from standard accretion thin-disc model \citep[e.g.,][]{2016ApJ...821...56F, 2018MNRAS.480.2881M}. In our analysis, we find the excess U-band lag (Figure~\ref{lag_versus_wavelength}), but with no corresponding excess U-band (Figure~\ref{log-log}). If we compare ICCF's best-fit disc normalization $\tau_{\rm 0}$ at $\lambda_{\rm 0}$ with CREAM, results in $\tau_{\rm 0}$ that is a factor of $1.6$ times larger, consistent with previous studies in which larger than expected values have been measured.
\subsubsection{Accretion disc spectrum}
\label{discussion_accretion_disc_spectrum}
Using CREAM's flux-flux method \citep{2017ApJ...835...65S}, we investigate the origin of the UV/Optical continuum variability. In Figure~\ref{flux-flux}, we observe linear relations with no significant curvature that would indicate a change in shape by the variable component across all the continuum light curves. This strong linear relationship implies that the variable component has a constant flux distribution \citep{1997MNRAS.292..273W}, consistent with a disc reprocessing model. The slopes $\Delta F_\nu$ of the best-fitting line of $f_\nu(\lambda, t)$ versus $X(t)$ give the spectrum of the variable component. We see that the spectral energy distribution (SED) of the variable spectrum resembles the $f_\nu \propto \lambda^{-1/3}$ spectrum expected for a steady-state blackbody accretion disc, after correcting for MW dust extinction. The intercepts $\bar{F}_\nu$ of the best-fitting line are used to derive the host-galaxy spectrum. We see that the variable component spectrum, after correcting for MW dust extinction, is very close to the $f_\nu\propto\nu^{1/3}$ spectrum predicted by standard thin-disc reprocessing model.\\
\indent Figure~\ref{predicted_spectra}, we draw the comparison between the observed mean accretion disc fluxes $f_\nu(\lambda)$ and a predicted blackbody accretion disc spectrum for a face-on disc, using CREAM fit estimates in Figure~\ref{temperature_profile}. Our finding of a similarly low disc surface brightness in 3C~120 as was found previously for NGC~5548 \citep{2017ApJ...835...65S}, suggests that this anomaly may be more widespread. We encourage similar analysis to measure both the disc flux and lag spectra in other targets, rather than focusing exclusively on the ICCF delay spectrum.

\subsection{Size-Luminosity Relation}
\begin{figure}
\centering
\includegraphics[width = \columnwidth]{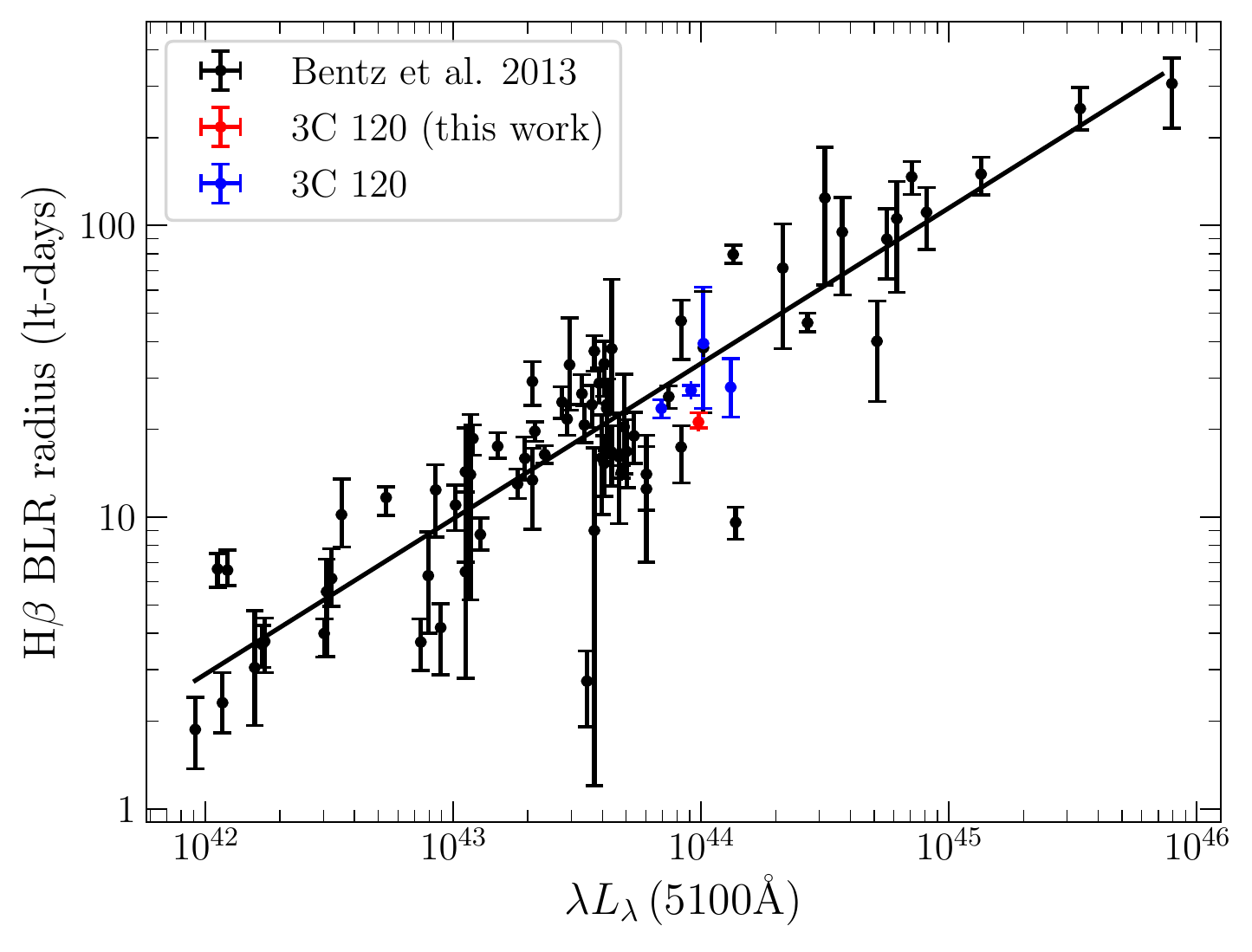}
\caption{Optical continuum luminosity and H$\beta$ lags for 3C~120 (this work) and other AGNs from \protect\cite{2013ApJ...767..149B}.}
\label{Rad_vs_lum}
\end{figure}

Figure~\ref{Rad_vs_lum} examines the location of our H$\beta$ lag measurement in the context of the $R_{\text{BLR}}-L$ relationship. We use CREAM's flux-flux analysis to determine the optical luminosity at 5100\AA~corresponding to our campaign, in a similar manner as, e.g., \cite{1997MNRAS.292..273W, 2012A&A...545A..84P}. This led to $\lambda L_\lambda(5100\AA)= 10^{43.99\pm0.01}$~erg~s$^{-1}$ after subtracting host-galaxy contribution estimated from Table~\ref{host_galaxy_subtracted_fluxes}. Figure~\ref{Rad_vs_lum} shows measurements of the size of the H$\beta$-emitting region as a function of the optical luminosity at $5100\text{\AA}$ using measurements from \cite{2013ApJ...767..149B}. Our measurement is consistent with the general H$\beta$ trend.
\subsection{$M_{\text{BH}}-\sigma_{*}$ Relation}
\begin{figure}
\centering
\includegraphics[width = \columnwidth]{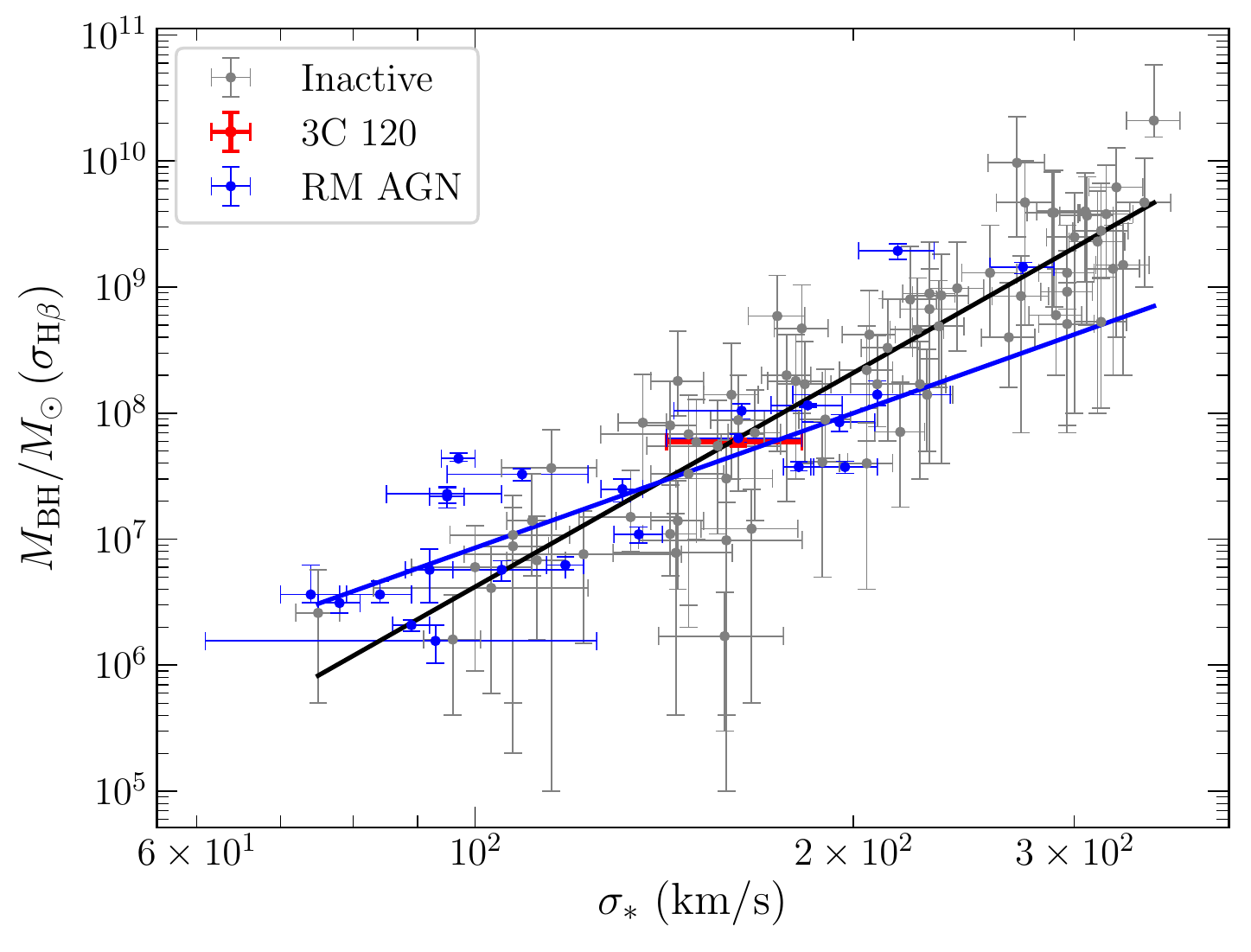}
\caption{$M_{\text{BH}}-\sigma_{*}$ relation of inactive galaxy sample (gray), compared with active galaxies with reverberation masses (blue). The reverberation masses were measured using a virial factor of $f = 5.2$. The solid lines are the best-fit lines to the reverberation sample ($\alpha = 8.00$ and $\beta = 3.55$) and inactive galaxy sample ($\alpha = 8.32$ and $\beta = 5.64$).}
\label{m_sigma}
\end{figure}

We also investigate if 3C~120 obeys the general trend of the scaling relations between $M_{\text{BH}}$ and host-galaxy properties. We adopt the stellar velocity dispersion of 3C~120 of $\sigma_{*} = 162\pm 20$ km s$^{-1}$ reported in \cite{1995ApJS...99...67N}. We provide $M_{\text{BH}}-\sigma_{*}$ relation combining the samples of \cite{2013ApJ...764..184M} for inactive galaxies, and the reverberation-mapped AGN sample of \cite{2010ApJ...716..269W}, along with our measurement of 3C~120 in Figure~\ref{m_sigma}. The best-fit lines are characterized by
\begin{equation}
\text{log}(M_{\text{BH}}/M_{\odot}) = \alpha + \beta\,\text{log}(\sigma_{*}/200\,\text{km s}^{-1})
\end{equation} 
For a sample of reverberation masses, \cite{2010ApJ...716..269W} measured $\alpha = 8.00\pm 0.24$ and $\beta = 3.55\pm 0.60$ assuming $\text{log}\,f = 0.72\pm 0.10$ with the intrinsic scatter of $\sigma_{\text{int}} = 0.43\pm 0.08$ dex. \cite{2013ApJ...764..184M} measured $\alpha = 8.32\pm 0.05$ and $\beta = 5.64\pm 0.32$ for a sample of 72 quiescent galaxies. In both cases, we find that our target shows no significant deviation from the $M_{\text{BH}}-\sigma_{*}$ relation for quiescent galaxies and reverberation based samples.\\
\section{Conclusion}
\label{conclusion}
We carried out fully robotic photometric and spectroscopic observations of the well-studied broad-line radio galaxy 3C~120 with the Las Cumbres Observatory (LCO) from 2016 December to 2018 April as part of the LCO AGN Key Project on Reverberation Mapping of Accretion Flows. We summarize the main results as follows:
\begin{itemize}
\item We used the ICCF method alongside CREAM to carry out multiple-line (H$\gamma$, He~II $\lambda 4686$, H$\beta$, He~I $\lambda 5876$) lag measurements in 3C~120. We find that the two techniques give consistent results to within uncertainties.
\item We find lags relative to V of 2.7~days for He~II and 15-18~days for He~I and Balmer lines, consistent with a decrease in ionisation with radius.
\item Combining the H$\beta$ lag ($21.2_{-1.0}^{+1.6}$ days) relative to V, and its velocity dispersion ($\sigma_{\rm line} = 1657\pm3$~km~s$^{-1}$) in the rms spectrum, we infer a central black hole mass ($6.3_{-0.3}^{+0.5}\times 10^{7}\,\text{M}_{\odot}$) that is in accord with results from previous studies after correcting to a common virial correction
factor $f=5.5$.
\item From our velocity-resolved H$\beta$ lag measurements, we deduced a symmetric pattern with a 25~day lag at line centre reducing to $\sim17$~days on both the blue and red wings of the line.
\item Our inter-band continuum variability study reveals wavelength-dependent delays with longer delays at longer wavelengths. The delays span 3 to 4 days, and are compatible with $\tau\propto\lambda^{4/3}$ apart from an excess of 1 or 2 days in the U-band.
\item Our CREAM analysis, fitting a face-on reverberating steady-state black body disc model, finds $M\,M_\odot= 10^{7.58\pm0.11}\,M_{\odot}^{2}/\text{yr}$ corresponding to $L/L_{\rm Edd}\sim0.4$ for $M=6.3\times10^7\,M_\odot$.
\item We measure the size of the disc to be 1.6 times larger than the prediction from CREAM thin-disc theory with $L/L_{\rm Edd}\sim0.4$, consistent with other RM continuum studies. 
\item The SED of the disc spectrum, isolated by variations, matches $F_{\nu}\propto\nu^{1/3}$ from disc theory with no sign of bound-free edges.
\end{itemize} 
\indent The CREAM thin-disc model generally fits the 3C~120 data relatively well. In the future, we want to explore departures from the thin-disc model by varying the inclination (from face-on), the power index $\alpha$ to see how it affects the mass accretion rate in 3C~120, and potentially the Eddington ratio in terms of the thin-disc theory. Due to the high quality of these observations, they lend themselves to potential further study for a more detailed model for 3C~120.

\section*{Acknowledgements}
MH and ERC acknowledge support from the South African National Research Foundation. KH acknowledges support from STFC grant ST/R000824/1. MH and SMC were supported in part by the Space Telescope Science Institute (STScI) Director's Discretionary Research Fund (DDRF). Research by AJB was supported by NSF grants AST-1412693 and AST-1907290. Research by DJS is supported by NSF grants AST-1821967, 1821987, 1813708, 1813466, and 1908972. TT acknowledges support by the NSF through grants AST-1412315 and AST-1907208 and by the Packard Foundation through a Packard Fellowship.

\section*{Data Availability}
The data underlying this article are available in the article and in its online supplementary material.


\bibliographystyle{mnras}
\bibliography{my_refs} 




\appendix

\section{Additional tables}
\begin{table*}
\caption{Continuum light curve data}
\label{continuum_lightcurve_data}
\begin{threeparttable}
\begin{tabular}{cccc}
\hline \hline
$\text{HJD}-2450000$ & Flux & Error &  Band\\
(days) & (mJy) & (mJy) &\\
\hline
7970.851 & 2.471 & 0.099 & U\\
7970.855 & 2.298 & 0.078 & U\\
7971.278 & 2.229 & 0.080 & U\\
7971.282 & 2.257 & 0.075 & U\\
7974.237 & 2.469 & 0.214 & U\\
\hline
7818.909 & 3.871 & 0.054 & $g'$\\
7818.911 & 3.736 & 0.051 & $g'$\\
7819.266 & 3.828 & 0.032 & $g'$\\
7819.268 & 3.752 & 0.031 & $g'$\\
7820.265 & 3.772 & 0.032 & $g'$\\
\hline
7751.044 & 5.322 & 0.017 & V\\
7751.048 & 5.285 & 0.014 & V\\
7756.114 & 5.226 & 0.013 & V\\
7756.118 & 5.206 & 0.013 & V\\
7760.362 & 5.184 & 0.017 & V\\
\hline
7818.924 & 7.177 & 0.078 & $r'$\\
7818.925 & 7.190 & 0.064 & $r'$\\
7821.276 & 7.288 & 0.046 & $r'$\\
7821.278 & 7.120 & 0.045 & $r'$\\
7822.269 & 7.272 & 0.046 & $r'$\\
\hline
7818.927 & 7.923 & 0.067 & $i'$\\
7818.929 & 7.846 & 0.067 & $i'$\\
7821.280 & 7.946 & 0.046 & $i'$\\
7821.282 & 7.890 & 0.047 & $i'$\\
7822.273 & 7.961 & 0.045 & $i'$\\
\hline
7818.932 & 9.356 & 0.069 & $z_s$\\
7818.935 & 9.544 & 0.066 & $z_s$\\
7819.971 & 9.448 & 0.120 & $z_s$\\
7819.974 & 9.393 & 0.181 & $z_s$\\
7821.284 & 9.518 & 0.049 & $z_s$\\
\hline
\end{tabular}
\begin{tablenotes}
\item[Note] The full table is available online. 
\end{tablenotes}
\end{threeparttable}
\end{table*}
\begin{table*}
\caption{Emission-line light curve data}
\label{continuum_lightcurve_data}
\begin{threeparttable}
\begin{tabular}{cccc}
\hline \hline
$\text{HJD}-2450000$ & Flux & Error &  Light Curve\\
(days) & $(10^{-12}\,\text{erg s}^{-1}\text{cm}^{-2})$ & $(10^{-12}\,\text{erg s}^{-1}\text{cm}^{-2})$ &\\
\hline
7823.763 & 1.524 & 0.026 & H$\gamma$\\
7823.785 & 1.503 & 0.030 & H$\gamma$\\
7827.749 & 1.579 & 0.021 & H$\gamma$\\
7827.770 & 1.538 & 0.022 & H$\gamma$\\
7831.756 & 1.596 & 0.022 & H$\gamma$\\
\hline
7823.763 & 0.360 & 0.031 & He~II $\lambda 4686$\\
7823.785 & 0.330 & 0.034 & He~II $\lambda 4686$\\
7827.749 & 0.203 & 0.025 & He~II $\lambda 4686$\\
7827.770 & 0.269 & 0.026 & He~II $\lambda 4686$\\
7831.756 & 0.313 & 0.026 & He~II $\lambda 4686$\\
\hline
7823.763 & 3.140 & 0.029 & H$\beta$ blue-side\\
7823.785 & 2.994 & 0.032 & H$\beta$ blue-side\\
7827.749 & 3.101 & 0.026 & H$\beta$ blue-side\\
7827.770 & 3.085 & 0.027 & H$\beta$ blue-side\\
7831.756 & 2.999 & 0.027 & H$\beta$ blue-side\\
\hline
7823.763 & 4.533 & 0.025 & H$\beta$ red-side\\
7823.785 & 4.458 & 0.027 & H$\beta$ red-side\\
7827.749 & 4.231 & 0.022 & H$\beta$ red-side\\
7827.770 & 4.224 & 0.024 & H$\beta$ red-side\\
7831.756 & 4.253 & 0.024 & H$\beta$ red-side\\
\hline
7823.763 & 1.087 & 0.016 & He~I $\lambda 5876$\\
7823.785 & 1.027 & 0.017 & He~I $\lambda 5876$\\
7827.749 & 0.961 & 0.014 & He~I $\lambda 5876$\\
7827.770 & 0.913 & 0.015 & He~I $\lambda 5876$\\
7831.756 & 0.899 & 0.015 & He~I $\lambda 5876$\\
\hline
\end{tabular}
\begin{tablenotes}
\item[Note] The full table is available online. 
\end{tablenotes}
\end{threeparttable}
\end{table*}
\bsp	
\label{lastpage}
\end{document}